\documentclass[preprint,onecolumn,aps,prd,groupedaddress,showpacs,nofootinbib]{revtex4}%

\usepackage{epsfig}
\usepackage{graphicx}
\usepackage{slashed}
\usepackage{amsmath}
\usepackage{bm}
\usepackage{relsize}
\RequirePackage{xspace}
\begin{document}


\title{\mbox{}\\[10pt]
Relativistic correction to color Octet \mbox{\boldmath $J/\psi$} production at
hadron colliders}

\author{Guang-Zhi Xu $^{(a,b)}$}
\email{ still200@gmail.com}
\author{ Yi-Jie Li $^{(a,b)}$}
\email{ yijiegood@gmail.com}
\author{Kui-Yong Liu $^{(b)}$}
\email{liukuiyong@lnu.edu.cn}
\author{Yu-Jie Zhang $^{(a)}$}
\email{nophy0@gmail.com}
\affiliation{ {\footnotesize (a)~School of Physics, Beihang University,
 Beijing 100191, China}\\
  {\footnotesize (b)~Department of Physics, Liaoning University, Shenyang 110036
, China}}


\begin{abstract}

The relativistic corrections to the color-octet  $J/\psi$
hadroproduction at the Tevatron and LHC are calculated up to
$\mathcal{O}(v^2)$ in nonrelativistic QCD  factorization frame.
The short
distance coefficients are obtained by matching full QCD with NRQCD
results for the partonic subprocess $g+g\to J/\psi ({}^1S_0^{[8]},{}^3S_1^{[8]},
{}^3P_J^{[8]})+g$, $q+\bar{q}\to J/\psi ({}^1S_0^{[8]},{}^3S_1^{[8]},
{}^3P_J^{[8]})+g$ and $g+q({\bar{q}})\to J/\psi ({}^1S_0^{[8]},{}^3S_1^{[8]},
{}^3P_J^{[8]})+q({\bar{q}})$.
The short distance coefficient ratios of relativistic correction to leading order
for color-octet states ${^1}S_{0}^{[8]}$, ${^3}S_{1}^{[8]}$, and ${^3}P_{J}^{[8]}$
at large $p_T$ are  approximately -5/6, -11/6, and -31/30, respectively, for each subprocess,
and it is 1/6 for color-singlet state  ${^3}S_{1}^{[1]}$.
If the higher order long distance matrix elements are estimated through velocity
scaling rule with adopting $v^2=0.23$ and the lower order long distance matrix elements are fixed,
the leading order cross sections of color-octet states  are reduced by about a factor of $20\sim40\%$ at large $p_T$ at
both the Tevatron and the LHC.
Comparing with QCD radiative corrections to color-octet states, relativistic correction is ignored along with $p_T$ increasing.
Using long distance matrix elements extracted from the fit to $J/\psi$ production at the Tevatron, we can find
the unpolarization cross sections of $J/\psi$ production at the LHC taking into account both QCD and relativistic corrections are changed by about $20\sim50\%$ of that considering only QCD corrections.
These results indicate that relativistic corrections may play an important
role  in $J/\psi$ production 
at the Tevatron and the LHC.

\end{abstract}


\maketitle

\section{Introduction}
Heavy quarkonium is an excellent candidate to probe quantum chromodynamics (QCD) from the high
energy to the low energy regimes. Nonrelativistic QCD (NRQCD) factorization
formalism was established\cite{Bodwin:1994jh} to describe the production and
decay  of heavy quarkonium. In the NRQCD approach,
the production and decay of heavy quarkonium is factored into short
distance coefficients and long distance matrix elements(LDMEs).
The short distance coefficients indicate the creation or annihilation of a
heavy quark pair
can be calculated perturbatively with the expansions
by the strong coupling constant $\alpha_s$. However, the LDMEs,  which
represent
the evolution of a free heavy quark pair into a bound state, can be
scaled by
the relative velocity $v$ between the quark and antiquark
and
obtained by lattice QCD or extracted from the experiment. $v^2$ is about $0.2\sim0.3$ for charmonium
and about $0.08\sim0.1$ for bottomonium. The color-octet
mechanism (COM)  was introduced here. The heavy quark pair should be a
color-singlet (CS) bound state at long distances, but it  may be in a
color-octet (CO) state at short distances.
NRQCD had achieved great success since it was proposed.  The COM was
applied to cancel the infrared divergences in the decay widths  of
$P$-wave \cite{Huang,Petrelli:1997} and $D$-wave\cite{He:2008xb,Fan:2009cj}
heavy quarkonium. However, difficulties were still encountered. The large discrepancy
between the experimental data and the theoretical calculation of $J/\psi$ and
$\psi^\prime$ unpolarization and polarization
production at Tevatron is an interesting phenomenon
that can verify NRQCD when
solved\cite{CDF:1992,arXiv:0704.0638}.
Theoretical prediction with COM contributions was introduced and was found to fit with the
experimental data on $J/\psi$ production  at Tevatron\cite{Braaten:1994}.
However, the CO contributions from gluon fragmentation
indicated that the $J/\psi$ was transversely polarized at large
$p_T$, which is   inconsistent with the
experimental data\cite{CDF:1992}.

The next-to-leading order (NLO) QCD corrections
and other possible solutions for $J/\psi$ hadroproduction were calculated to resolve the
$J/\psi$
hadronic production and polarization puzzle\cite{Campbell:2007,Gong:2008}.
The calculation enhanced the CS cross sections at
large $p_T$ by approximately an order of magnitude. However,
the large discrepancy between the CS predictions and
experimental data remains unsolved. The relativistic correction to CS
$J/\psi$ hadroproduction was insignificant\cite{Fan:2009zq}. The NLO QCD corrections of COM $J/\psi$ hadroproduction were also calculated to formulate a possible solution to the
long-standing $J/\psi$ polarization puzzle
\cite{Chao:2012iv,Ma:2010jj,Butenschoen:2011yh}.
The spin-flip interactions in the spin density matrix of the hardronization of a color-octet charm quark pair had been examined in Ref.\cite{Liu:2006hc}.
A similar large discrepancy was found in double-charmonium production  at $B$
factories\cite{Abe:2002rb,BaBar:2005,Braaten:2002fi}.
A great deal of work had been
performed on this area, and these discrepancies can apparently  be resolved by
including NLO QCD
corrections\cite{Zhang:2005cha,Zhang:2006ay,Gong:2007db,zhangma08}
and relativistic corrections\cite{Bodwin:2006dm,He:2007te,Jia:2009np,He:2009uf}.
The data from $B$ factories  highlight  that the
COM LDMEs of $J/\psi$ production may
be smaller than previously expected\cite{Ma:2008gq,Gong:2009kp,Jia:2009np,
He:2009uf,Zhang:2009ym}.
Relativistic corrections have also been studied in Ref.\cite{Huang:1996bk} for heavy quarkonium decay, in Ref.\cite{Paranavitane:2000if} for $J/\psi$ photoproduction, in Ref.\cite{Ma:2000qn} for $J/\psi$ production in $b$ decay, and in Ref.\cite{Bodwin:2003wh} for  gluon fragmentation into spin triplet $S$ wave
quarkonium.
More information about heavy quarkonium physics can be found in
Refs.\cite{Kramer:2001,Lansberg:2006dh,Brambilla:2010cs}.

In this paper, the effects of relativistic
corrections to the COM $J/\psi$ hadroproduction at Tevatron and
LHC were estimated based on NRQCD. The short distance coefficients
were calculated up to
$\mathcal{O}(v^2)$.
Many free LDMEs were realized at $\mathcal{O}(v^2)$, which were estimated
according to the velocity scaling rules of
NRQCD with $v^2=0.23$\cite{velocity}.

The paper is organized as follows. In Sec. II, the frame of calculation
is introduced for  the relativistic correction of both the $S$- and $P$-wave
states in NRQCD frame.
Section III provides the numerical result. Finally, a brief
summary of this work is presented.
\newline

\section{Relativistic Corrections of Cross Section in NRQCD}
\label{sec:two}
We only consider $J/\psi$ direct production at high energy hadron colliders, which contributes $70\%$ to the prompt cross section. The differential cross section of direct production can be obtained by integrating the cross sections of parton level as the following expression:
\begin{eqnarray}
d\sigma\big(p+p(\bar{p}){\rightarrow}J/\psi+X\big)
=\sum_{a,b,d}{\int}dx_1dx_2{f_{a/p}(x_1)}{f_{b/p(\bar{p})}(x_2)}
d\hat{\sigma}(a+b{\rightarrow}J/\psi+d).
\end{eqnarray}
where $f_{a(b)/p(\bar{p})}(x_i)$ is the parton distribution function(PDF), and $x_i$ is the parton momentum fraction denoted the fraction parton carried from proton or antiproton. The sum is over all the partonic subprocesses including
\begin{eqnarray}
&&g+g{\rightarrow}J/\psi+g\nonumber\\
&&g+q(\bar{q}){\rightarrow}J/\psi+q(\bar{q})\nonumber\\
&&q+\bar{q}{\rightarrow}J/\psi+g\nonumber.
\end{eqnarray}

As shown at the beginning of this paper, under the NRQCD frame, the computation to cross section of
each
subprocess can be divided into two parts: short distance coefficients and LDMEs:
\begin{equation}
\label{factorization} d\hat{\sigma}(a(k_1)+b(k_2){\rightarrow}J/\psi(P)+d(k_3))=\sum_n\frac{F_n(ab)}{m_c^{d_n-4}}\langle0|\mathcal{O}_n^{J/\psi}|0\rangle.
\end{equation}
On the right-hand side of the equation, the cross section is expanded to sensible
Fock states noted by the subscript $n$. $F_n$, i.e.,  short distance coefficients,
which describe the process that  produces
intermediate $Q\bar{Q}$ in a short range before
heavy quark and antiquark hadronization to the physical meson state.
Here we use initial partons to mark the short distance coefficients for different subprocesses.
$\langle0|\mathcal{O}_n^{J/\psi}|0\rangle$
are the long distance matrix elements that represent the hadronization $Q\bar{Q}$
evolutes to the CS final state by emitting soft gluons. $\mathcal{O}_n^{J/\psi}$ are
local four fermion operators. The factor of $m_c^{d_n-4}$ is introduced to make
$F_n$ dimensionless.

In this section, our calculation on the differential cross section for this
process in the NRQCD factorization formula is divided into three parts, namely,
kinematics, long distance matrix elements,  and short distance coefficients.

\subsection{Kinematics}

We denote the three relative momenta between heavy quark and antiquark as
$2\vec{q}$, with $|\vec{q}|\sim{m_{c}v}$, in $J/\psi$ rest frame, where $m_c$
is
the mass of charm quark and $v$ is the three relative velocity of
quark or antiquark in this frame. Thus, the momenta for the quark and antiquark are expressed as\cite{He:2007te,Ma:2000,fourmomenta}
\begin{eqnarray}
p_c&=&(E_q, \vec{q}),\nonumber \\
p_{\bar{c}}&=&(E_q, -\vec{q}).
\end{eqnarray}
where $E_q=\sqrt{m_{c}^2+|\vec{q}|^2}$ is the rest energy of both the quark and
antiquark, and  $2E_q$ is the invariable mass of $J/\psi$. When boosting to an
arbitrary frame,
\begin{eqnarray}
\begin{array}{l}
p_c\rightarrow\frac{1}{2}P+q,\quad
p_{\bar{c}}\rightarrow\frac{1}{2}P-q.
\end{array}
\end{eqnarray}
where $P$ is the four momenta of $J/\psi$, and $q$ receives
the boost from $(0, \vec{q})$.

The Lorentz invariant Mandelstam variables are defined as
\begin{equation}
s=(k_1+k_2)^2=(P+k_3)^2, \nonumber\\
\end{equation}
\begin{equation}
t=(k_1-P)^2=(k_2-k_3)^2, \nonumber\\
\end{equation}
\begin{equation}
u=(k_1-k_3)^2=(k_2-P)^2.\nonumber
\end{equation}
with the relationship $s+t+u=P^2=4E_q^2$. Here, s is $|\vec{q}|^2$ independence. To expand $t$, $u$ in terms of $E_q(i.e. |\vec{q}|^2)$, we can first write down $t$, $u$ in the center of initial partons mass frame:
\begin{eqnarray}\label{eq:tuexpand}
t(|\vec{q}|)&=&-(s-4E_q^2)(1-cos\theta)/2=\frac{s-4E_q^2}{s-4m_c^2}t(0),\nonumber\\
u(|\vec{q}|)&=&-(s-4E_q^2)(1+cos\theta)/2=\frac{s-4E_q^2}{s-4m_c^2}u(0),
\end{eqnarray}
where $t(0) ,u(0)$ are Lorentz invariants of $|\vec{q}|^2$ independence and satisfies $s+t(0)+u(0)=4m_c^2$. These relations between $t(|\vec{q}|)\big(u(|\vec{q}|)\big)$ and $t(0)\big(u(0)\big)$ are also satisfied when boosting to arbitrary frame.
In our subsequent calculation and result,
we adopt $t$($u$) to represent $t(0)$($u(0)$) directly for simplification.

The {\tt FeynArts}~\cite{feynarts} package was used  to generate Feynman diagrams
and amplitudes, and the {\tt FeynCalc}~\cite{Mertig:an}  package was used to handle
amplitudes. The numerical  phase space was integrated  with Fortran.

\subsection{ Long Distance Matrix Elements}
According to NRQCD factorization, the differential cross section of
each partonic subprocess
up to next order in $v^2$ to CS state $^3S_1^{[1]}$ and CO states
${^1}S_{0}^{[8]}$, ${^3}S_{1}^{[8]}$,  ${^3}P_{J}^{[8]}$, can be expressed as
\begin{eqnarray}\label{eq:dsigma}
d\sigma&=&d\sigma_{lo}[^3S_1^{[1]}]+d\sigma_{lo}[^1S_0^{[8]}]+d\sigma_{lo}
[^3S_1^{[8]}]+d\sigma_{lo}[^3P_J^{[8]}]
\nonumber\\
&+&d\sigma_{rc}[^3S_1^{[1]}]+d\sigma_{rc}[^1S_0^{[8]}]+d\sigma_{rc}[^3S_1^{[8]}]
+d\sigma_{rc}[^3P_J^{[8]}].
\end{eqnarray}
In this expression,  relativistic correction parts, denoted as ''$rc$'',
can easily be distinguished from LO, denoted as ''$lo$''. Ref.$[1]$ corresponds to CS, and Ref.$[8]$  corresponds to CO.  In addition,
each differential cross section to different Fock states should be
divided in short distance coefficient part and LDMEs.
We can
introduce $F(^{2s+1}L_J^{[c]})$ to express the short distance coefficient of the LO cross
section, corresponding to $G(^{2s+1}L_J^{[c]})$ for relativistic correction.
Many LDMEs are
presented
, all of which are
denoted by $\langle0|\mathcal{O}^{J/\psi}(^{2s+1}L_J^{[c]})|0\rangle$
and $\langle0|\mathcal{P}^{J/\psi}(^{2s+1}L_J^{[c]})|0\rangle$ for the LO and relativistic
correction term respectively. The explicit expressions of the ten
four-fermion operators are\cite{Bodwin:1994jh}
\begin{eqnarray}
<0|\mathcal{O}^{J/\psi}({}^3S_1^{[1]})|0>&=&<0|\chi^\dagger\sigma^{i}\psi
(a^\dagger_{\psi}a_{\psi})
\psi^\dagger\sigma^i\chi|0>,
\nonumber\\
<0|\mathcal{P}^{J/\psi}({}^3S_1^{[1]})|0>&=&<0|\frac{1}{2}[\chi^\dagger\sigma^{i}\psi
(a^\dagger_{\psi}a_{\psi})\psi^\dagger\sigma^i(-\frac{i}{2}
\overleftrightarrow{\mathbf{D}})^2\chi+h.c.]|0>,
\nonumber\\
<0|\mathcal{O}^{J/\psi}({}^1S_0^{[8]})|0>&=&<0|\chi^\dagger{T^a}\psi
(a^\dagger_{\psi}a_{\psi})
\psi^\dagger{T^a}\chi|0>,
\nonumber\\
<0|\mathcal{P}^{J/\psi}({}^1S_0^{[8]})|0>&=&<0|\frac{1}{2}[\chi^\dagger{T^a}\psi
(a^\dagger_{\psi}a_{\psi})\psi^\dagger{T^a}(-\frac{i}{2}\overleftrightarrow
{\mathbf{D}})^2\chi+h.c.]|0>,
\nonumber\\
<0|\mathcal{O}^{J/\psi}({}^3S_1^{[8]})|0>&=&<0|\chi^\dagger{T^a}\sigma^{i}\psi
(a^\dagger_{\psi}a_{\psi})
\psi^\dagger{T^a}\sigma^i\chi|0>,
\nonumber\\
<0|\mathcal{P}^{J/\psi}({}^3S_1^{[8]})|0>&=&<0|\frac{1}{2}[\chi^\dagger{T^a}\sigma^{i}\psi
(a^\dagger_{\psi}a_{\psi})\psi^\dagger{T^a}\sigma^i(-\frac{i}{2}
\overleftrightarrow{\mathbf{D}})^2\chi+h.c.]|0>,
\nonumber\\
<0|\mathcal{O}^{J/\psi}({}^3P_0^{[8]})|0>&=&{1 \over 3} <0|\chi^\dagger{T^a}(-\frac{i}{2}
\overleftrightarrow{D} \cdot \sigma)\psi(a^\dagger_{\psi}a_{\psi})
\psi^\dagger{T^a}(-\frac{i}{2}\overleftrightarrow{D}\cdot {\sigma})\chi|0>,
\nonumber\\
<0|\mathcal{O}^{J/\psi}({}^3P_1^{[8]})|0>&=&{1 \over 2} <0|\chi^\dagger{T^a}(-\frac{i}{2}
\overleftrightarrow{D}\times \sigma)\psi(a^\dagger_{\psi}a_{\psi})
\psi^\dagger{T^a}(-\frac{i}{2}\overleftrightarrow{D}\times \sigma)\chi|0>,
\nonumber\\
<0|\mathcal{O}^{J/\psi}({}^3P_2^{[8]})|0>&=&<0|\chi^\dagger{T^a}(-\frac{i}{2}
\overleftrightarrow{D^{(i}}\sigma^{j)})\psi(a^\dagger_{\psi}a_{\psi})
\psi^\dagger{T^a}(-\frac{i}{2}\overleftrightarrow{D^{(i}}\sigma^{j)})\chi|0>,
\nonumber\\
<0|\mathcal{P}^{J/\psi}({}^3P_J^{[8]})|0>&=&<0|\frac{1}{2}[\chi^\dagger{T^a}
(-\frac{i}{2}\overleftrightarrow{D^i}\sigma^j)\psi
(a^\dagger_{\psi}a_{\psi})\psi^\dagger{T^a}(-\frac{i}{2}
\overleftrightarrow{\mathbf{D}})^2(-\frac{i}{2}\overleftrightarrow{D^i}
\sigma^j)\chi+h.c.]|0>,
\end{eqnarray}
where $\chi$ and $\psi$ are the Pauli spinors describing anticharm quark
creation and charm quark annihilation, respectively. $T$ is the $SU(3)$ color matrix.
$\sigma$ is the Pauli
matrices and $\mathbf{D}$ is the gauge-covariant derivative with
$\overleftrightarrow{\mathbf{D}}=\overrightarrow{\mathbf{D}}-
\overleftarrow{\mathbf{D}}$.  $\overleftrightarrow{D^{(i}}\sigma^{j)}$ is used as   the notation
for the symmetric traceless component of a tensor: $\overleftrightarrow{D^{(i}}\sigma^{j)}=
(\overleftrightarrow{D^{i}}\sigma^{j}+\overleftrightarrow{D^{i}}\sigma^{j})/2-
\overleftrightarrow{D^{k}}\sigma^{k}\delta^{ij}/3$. Here we have
\begin{equation}\label{eq:8ME}
v^2=\frac{\langle0|\mathcal{P}^{J/\psi}(^{2s+1}L_J^{[c]})|0\rangle}
{m_c^2\langle0|\mathcal{O}^{J/\psi}(^{2s+1}L_J^{[c]})|0\rangle}.
\end{equation}
It should be noted that
\begin{eqnarray}
<0|\mathcal{O}^{J/\psi}({}^3P_J^{[8]})|0>&=&
(2J+1)(1+\mathcal{O}(v^2))<0|\mathcal{O}^{J/\psi}({}^3P_0^{[8]})|0>, \nonumber \\
<0|\mathcal{P}^{J/\psi}({}^3P_J^{[8]})|0>&=&
(2J+1)(1+\mathcal{O}(v^2))<0|\mathcal{P}^{J/\psi}({}^3P_0^{[8]})|0>\\
&\sim& {\cal O}(v^2) <0|\mathcal{O}^{J/\psi}({}^3P_J^{[8]})|0>\nonumber .
\end{eqnarray}
To NLO in $v^2$, we can ignore ${\cal O}(v^4)$ terms and set
\begin{eqnarray}
<0|\mathcal{P}^{J/\psi}({}^3P_J^{[8]})|0>&=&
(2J+1)<0|\mathcal{P}^{J/\psi}({}^3P_0^{[8]})|0> .\nonumber
\end{eqnarray}
So there are four CO LDMEs for $P$-wave, four CO LDMEs for $S$-wave and two CS LDMEs  at NLO in $v^2$. The LDMEs of heavy quarkonium decay may be determined by potential
model\cite{Bodwin:2007fz,Bodwin:2006dm}, lattice
calculations\cite{Bodwin:1996tg}, or phenomenological extraction
from experimental data\cite{Fan:2009zq,Guo:2011tz}. But it is very difficult to determine the production of CO LDMEs. Recently, two groups fitted CO LDMEs $<0|\mathcal{O}^{J/\psi}(^{2s+1}L_J^{[8]})|0>$  to NLO in $\alpha_s$. It is
\begin{eqnarray}
<0|\mathcal{O}^{J/\psi}({}^1S_0^{[8]})|0>&=&(8.90\pm0.98)\times10^{-2}~GeV^3 , \nonumber \\
<0|\mathcal{O}^{J/\psi}({}^3S_1^{[8]})|0>&=&(0.3\pm0.12)\times10^{-3}~GeV^3 , \nonumber \\<0|\mathcal{O}^{J/\psi}({}^3P_0^{[8]})|0>/m_c^2&=&(0.56\pm0.21)\times10^{-2}~GeV^3, \end{eqnarray}
with data of $J/\psi$ production and polarization at $p_t> 7~ GeV$ at Tevatron in Ref.\cite{Ma:2010jj} and
\begin{eqnarray}
<0|\mathcal{O}^{J/\psi}({}^1S_0^{[8]})|0>&=&(4.50\pm0.72)\times10^{-2}~GeV^3  , \nonumber \\
<0|\mathcal{O}^{J/\psi}({}^3S_1^{[8]})|0>&=&(3.12\pm0.93)\times10^{-3}~GeV^3 , \nonumber \\<0|\mathcal{O}^{J/\psi}({}^3P_0^{[8]})|0>&=&(-1.21\pm0.35)\times10^{-2}~GeV^5 ,\end{eqnarray}
with data of $J/\psi$ production at $p_t> 3~ GeV$  at Tevatron and  $p_T>2.5~GeV$ at HERA in Ref.\cite{Butenschoen:2011yh}. The two series CO LDMEs are not consistent with each other. For the three CO $P$ wave LDMEs $<0|\mathcal{O}^{J/\psi}({}^3P_J^{[8]})|0>$, it is hard to determine.
To simplify the discussion of the
numerical result, it is assumed that
\begin{eqnarray}
<0|\mathcal{O}^{J/\psi}({}^3P_J^{[8]})|0>&=&
(2J+1)<0|\mathcal{O}^{J/\psi}({}^3P_0^{[8]})|0> . \end{eqnarray}
At the same time,
we can estimate the relation between their order from the Gremm-Kapustin relation \cite{Gremm:1997dq} in the weak-coupling
regime
\begin{equation}
v^2=v_1^2=v_8^2=\frac{M_{J/\psi}-2m_c^{pole}}
{2m_c^{QCD}},
\end{equation}
where  $m_c^{QCD}$ is the mass of charm quark that appears in the NRQCD actions and $m_c^{pole}$ is the pole mass of charm quark. This equation was given only for CS in Ref.\cite{Gremm:1997dq}. This is the same with Ref.\cite{Bodwin:2003wh}, and we can get $v_1^2=v_8^2$.
If we select $M_{J/\psi}=3.1~GeV$ and
$m_c^{QCD}=m_c^{pole}=1.39~GeV$
, we can get $v^2\sim0.23$.

After those presses, there are three CO LDMEs in the numerical calculation.

\subsection{Short distance coefficients calculation}

The short distance coefficients can be evaluated by matching the computations
of perturbative QCD and NRQCD:
\begin{eqnarray}
d\sigma\Big|_{pert~QCD}
&&=\sum_n\frac{F_n}{m_c^{d_n-4}}\langle0|\mathcal{O}_n^{c\bar{c}}|0
\rangle\Big|_{pert~NRQCD}.
\end{eqnarray}
The covariant projection operator method should be adopted to
compute the expression on the left-hand side of the equation.
Using this method, spin-singlet and spin-triplet combinations of
spinor bilinears in the amplitudes can be written in covariant form.
For the spin-singlet case,
\begin{eqnarray}
&&\sum_{s\bar{s}}v(s)\bar{u}(\bar{s})\langle\frac{1}{2}, s;\frac{1}{2},
\bar{s}|0, 0\rangle\nonumber\\
&&=\frac{1}{2\sqrt{2}(E_q+m)}(-\slashed{p}_{\bar{c}}+m_c)\gamma_5
\frac{\slashed{P}+2E_q}{2E_q}(\slashed{p}_c+m_c).
\end{eqnarray}
For spin-triplet case, the expression is defined as
\begin{eqnarray}
&&\sum_{s\bar{s}}v(s)\bar{u}(\bar{s})\langle\frac{1}{2}, s;\frac{1}{2},
\bar{s}|1, S_z\rangle\nonumber\\
&&=\frac{1}{2\sqrt{2}(E_q+m)}(-\slashed{p}_{\bar{c}}+m_c)\slashed{\epsilon}
\frac{\slashed{P}+2E_q}{2E_q}(\slashed{p}_c+m_c),
\end{eqnarray}
where $\epsilon$ denotes the polarization vector of the spin-triplet state.
In our calculation, Dirac spinors are normalized as $\bar{u}u=-\bar{v}v=2m_c$.

The differential cross section of each state then satisfies:
\begin{eqnarray}
d\sigma(^{(2s+1)}L_J^{[c]}){\sim}\bar{\sum}|\mathcal{M}(a+b{\rightarrow}
(c\bar{c})(^{(2s+1)}L_J^{[c]})+d)|^2\langle0|\mathcal{O}^{J/\psi}(^{2s+1}L_J^{[c]})|0\rangle,
\end{eqnarray}
where $\bar{\sum}$ means sum over the final state color and
polarization and average over initial states.
According to this expression and Eq.(\ref{eq:8ME}),
expanding the cross section to next leading order of $v^2$ is to expand the amplitude
squared on the right side of the above expression to $\mathcal{O}(|\vec{q}|^2)$.

Next, we prepare to expand the short distance coefficients to the next order in $v^2$.
First, we expand each Fock state amplitude, including the $S$-wave and $P$-wave states,
in
terms
of the relative momentum $|\vec{q}|$:
\begin{eqnarray}
&&\mathcal{M}(a+b{\rightarrow}(c\bar{c})(^3S_1^{[1, 8]})+d)\nonumber\\
&&=\epsilon_{\rho}(\mathcal{M}^{\rho}_{t}\Big|_{q=0}+\frac{1}{2}q^{\alpha}q^{\beta}
\frac{\partial^2(\sqrt{\frac{m_c}{E_q}}\mathcal{M}^{\rho}_{t})}{\partial
q^{\alpha}\partial
q^{\beta}}\Big|_{q=0})+\mathcal{O}(q^4).
\end{eqnarray}
\begin{eqnarray}
&&\mathcal{M}(a+b{\rightarrow}(c\bar{c})(^1S_0^{[8]})+d)\nonumber\\
&&=\mathcal{M}_{s}\Big|_{q=0}+\frac{1}{2}q^{\alpha}q^{\beta}
\frac{\partial^2(\sqrt{\frac{m_c}{E_q}}\mathcal{M}_{s})}{\partial q^{\alpha}\partial
q^{\beta}}\Big|_{q=0}+\mathcal{O}(q^4).
\end{eqnarray}
\begin{eqnarray}
&&\mathcal{M}(a+b{\rightarrow}(c\bar{c})(^3P_J^{[8]})+d)=
\epsilon_{\rho}(s_z)\epsilon_{\sigma}(L_z)(\frac{\partial
\mathcal{M}^{\rho}_{t}}{\partial{q^{\sigma}}}\Big|_{q=0}\nonumber\\
&&+\frac{1}{6}q^{\alpha}q^{\beta}
\frac{\partial^3(\sqrt{\frac{m_c}{E_q}}\mathcal{M}^{\rho}_{t})}
{\partial q^{\alpha}\partial
q^{\beta}\partial{q^\sigma}}\Big|_{q=0})+\mathcal{O}(q^4).
\end{eqnarray}
The factor $\sqrt{\frac{m_c}{E_q}}$ comes from the relativistic
normalization of $c\bar{c}$ state. Odd power terms of four-momentum
$q$ vanish in either the $S$-wave or the $P$-wave amplitudes,
where $\mathcal{M}_{t}$ and $\mathcal{M}_{s}$ are inclusive production
amplitudes to triplet and singlet $c\bar{c}$, respectively.
\begin{equation}
\mathcal{M}_{s}=\sum_{s\bar{s}}\sum_{ij}\langle\frac{1}{2}, s;\frac{1}{2},
\bar{s}|0, 0\rangle\langle3i;\bar{3j}|1, 8a\rangle\mathcal{A}
(a+b{\rightarrow}c^i+\bar{c}^j+d).\nonumber
\end{equation}
\begin{equation}
\mathcal{M}_{t}=\sum_{s\bar{s}}\sum_{ij}\langle\frac{1}{2}, s;
\frac{1}{2}, \bar{s}|1, S_z\rangle\langle3i;\bar{3j}|1, 8a\rangle\mathcal{A}
(a+b{\rightarrow}c^i+\bar{c}^j+d).\nonumber
\end{equation}
In evaluating the amplitudes in power series in $|\vec{q}|$, it needs to be integrated over the space angle to $\vec{q}$. We can obtain the following replacements to extract the contribution of fixed power of $|\vec{q}|$:

For $S$-wave case:
\begin{equation}
q^{\alpha}q^{\beta}{\rightarrow}\frac{1}{3}\lvert\vec{q}\rvert{^2}\Pi^{\alpha\beta}.
\end{equation}

For $P$-wave case:
\begin{equation}
q^{\alpha}
q^{\beta}q^{\sigma}{\rightarrow}\frac{1}{5}\lvert\vec{q}\rvert{^3}\big[\Pi^{\alpha\beta}\epsilon^{\sigma}(L_z)
+\Pi^{\alpha\sigma}\epsilon^{\beta}(L_z)+\Pi^{\beta\sigma}\epsilon^{\alpha}(L_z)\big],
\end{equation}
where $\Pi^{\mu\nu}=-g^{\mu\nu}+\frac{P^{\mu}P^{\nu}}{P^2}$ and $\epsilon(L_z)$ is the orbital polarization vector of $P$-wave states. Subsequently, by multiplying the complex conjugate of the amplitude, the amplitude squared up to the next order can be obtained:
\begin{eqnarray}\label{eq:ampSquare3s18}
\sum|\mathcal{M}({}^3S_1^{[1,8]})|^2&=&\mathcal{M}_t^{\rho}(0)
\mathcal{M}_t^{{\lambda}*}(0)\sum_{s_z}\epsilon_{\rho}\epsilon^*_{\lambda}
\nonumber\\
&+&\frac{1}{3}|\vec{q}|^2\left[\left(\Pi^{\alpha\beta}\frac{\partial^2
(\sqrt{\frac{m_c}{E_q}}\mathcal{M}_t^{\rho})}{\partial
q^{\alpha}\partial q^{\beta}}\right)_{q=0}\mathcal{M}_t^{*{\lambda}}(0)
\right](\sum_{s_z}\epsilon_{\rho}\epsilon^*_{\lambda})_{q=0}+\mathcal{O}(v^4).
\end{eqnarray}
\begin{eqnarray}\label{eq:ampSquare1s08}
\sum|\mathcal{M}({}^1S_0^{[8]})|^2&=&\mathcal{M}_s(0)\mathcal{M}_s^{*}(0)
+\frac{1}{3}|\vec{q}|^2\left[\left(\Pi^{\alpha\beta}\frac{\partial^2
(\sqrt{\frac{m_c}{E_q}}\mathcal{M}_s)}{\partial
q^{\alpha}\partial q^{\beta}}\right)_{q=0}\mathcal{M}_s^{*}(0)
\right]+\mathcal{O}(v^4).
\end{eqnarray}
\begin{eqnarray}\label{eq:ampSquare3pj8}
\sum|\mathcal{M}({}^3P_J^{[8]})|^2&=&|\vec{q}|^2
\frac{\partial\mathcal{M}_t^{\rho}}{\partial{q^\alpha}}\Big|_{q=0}
\frac{\partial\mathcal{M}_t^{*{\lambda}}}{\partial{q^\beta}}
\Big|_{q=0}\sum_{L_z}\epsilon_{\alpha}\epsilon^*_{\beta}\sum_{s_z}\epsilon_{\rho}\epsilon^*_{\lambda}
\nonumber\\
&+&\frac{1}{15}|\vec{q}|^4\bigg[\left(\Pi^{\sigma\tau}
(\frac{\partial^3}{\partial{q^\alpha}\partial{q^\sigma}
\partial{q^\tau}}+\frac{\partial^3}{\partial{q^\sigma}
\partial{q^\alpha}\partial{q^\tau}}+\frac{\partial^3}
{\partial{q^\tau}\partial{q^\sigma}\partial{q^\alpha}})
(\sqrt{\frac{m_c}{E_q}}\mathcal{M}_t^{\rho})\right)\times
\nonumber\\
&&\frac{\partial\mathcal{M}_t^{*{\lambda}}}{\partial{q^\beta}}
(\sum_{L_z}\epsilon_{\alpha}\epsilon^*_{\beta})(\sum_{s_z}\epsilon_{\rho}\epsilon^*_{\lambda})
\bigg]_{q=0}+\mathcal{O}(v^6).
\end{eqnarray}
Any term, which is in the order
of $|\vec{q}|^2$, must not be missed to
obtain the correction up to the order of $v^2$.
In the three expressions above, the
first term on the right side of each equation can be
expressed in terms of kinematics variables $s,t(|\vec{q}|),u(|\vec{q}|)$.
Here $t(|\vec{q}|),u(|\vec{q}|)$ is $|\vec{q}|$ dependence and should be expanded by Eq.(\ref{eq:tuexpand}).
 The sum of terms in the order
of $|\vec{q}|^2$ in the first term as well as all the
second term is the contribution of the next leading order.
Orbit polarization sum $\sum_{L_z}$ and  spin-triplet polarization sum $\sum_{s_z}$
are equal to $\Pi^{\rho\lambda}(\Pi^{\alpha\beta})$.
According to the expression of $\Pi$ mentioned above, the $q$ dependence of $\Pi$ only appears in the denominator $P^2$ which equals to $4E_q^2$ and only contains even powers of four momentum $q$. So in the computation of unpolarized cross section to next order of $v^2$ as in Eqs.(\ref{eq:ampSquare3s18},\ref{eq:ampSquare3pj8}), expanding the polarization vector in order of $v^2$ is to handle the sum expression $\Pi$.

Therefore,
the differential cross section in Eq.(\ref{eq:dsigma}) takes the following form:
\begin{eqnarray}\label{crosssection}
d\hat{\sigma}(a+b \rightarrow
J/\psi+d)&=&\Bigg(
\frac{F({}^3S_1^{[1]})}{m_c^2}\langle0|\mathcal{O}^{J/\psi}({}^3S_1^{[1]})|0\rangle
+\frac{G({}^3S_1^{[1]})}{m_c^4}\langle0|\mathcal{P}^{J/\psi}({}^3S_1^{[1]})|0\rangle
+
\nonumber \\
&&
\frac{F({}^1S_0^{[8]})}{m_c^2}\langle0|\mathcal{O}^{J/\psi}({}^1S_0^{[8]})|0\rangle
+\frac{G({}^1S_0^{[8]})}{m_c^4}\langle0|\mathcal{P}^{J/\psi}({}^1S_0^{[8]})|0\rangle
+
\nonumber \\
&&
\frac{F({}^3S_1^{[8]})}{m_c^2}\langle0|\mathcal{O}^{J/\psi}({}^3S_1^{[8]})|0\rangle
+\frac{G({}^3S_1^{[8]})}{m_c^4}\langle0|\mathcal{P}^{J/\psi}({}^3S_1^{[8]})|0\rangle
+
\nonumber \\
&&\frac{F({}^3P_0^{[8]})}{m_c^2}\langle0|\mathcal{O}^{J/\psi}({}^3P_0^{[8]})|0\rangle
+\frac{G({}^3P_0^{[8]})}{m_c^4}\langle0|\mathcal{P}^{J/\psi}({}^3P_0^{[8]})|0\rangle
\Bigg)\times\nonumber \\
&&\Big(1
+\mathcal{O}(v^4)\Big).
\end{eqnarray}
The explicit expressions of the short distance
coefficients
to the relativistic correction of CO states ${^1}S_{0}^{[8]}$ and
${^3}S_{1}^{[8]}$ , ${^3}P_{J}^{[8]}$ for partonic processes $gg{\rightarrow}{J/\psi}g$, $gq(\bar{q}){\rightarrow}{J/\psi}q(\bar{q})$ and $q\bar{q}{\rightarrow}{J/\psi}g$
are relegated to the Appendix.
The result of our relativistic correction of $^3S_1^{[1]}$ is
consistent with that of Ref.\cite{Fan:2009zq} and was not given in this paper.

\section{numerical result and discussion}

\begin{figure}[h]
\begin{center}
\hspace{-1.4cm}\includegraphics[width=0.4\textwidth]{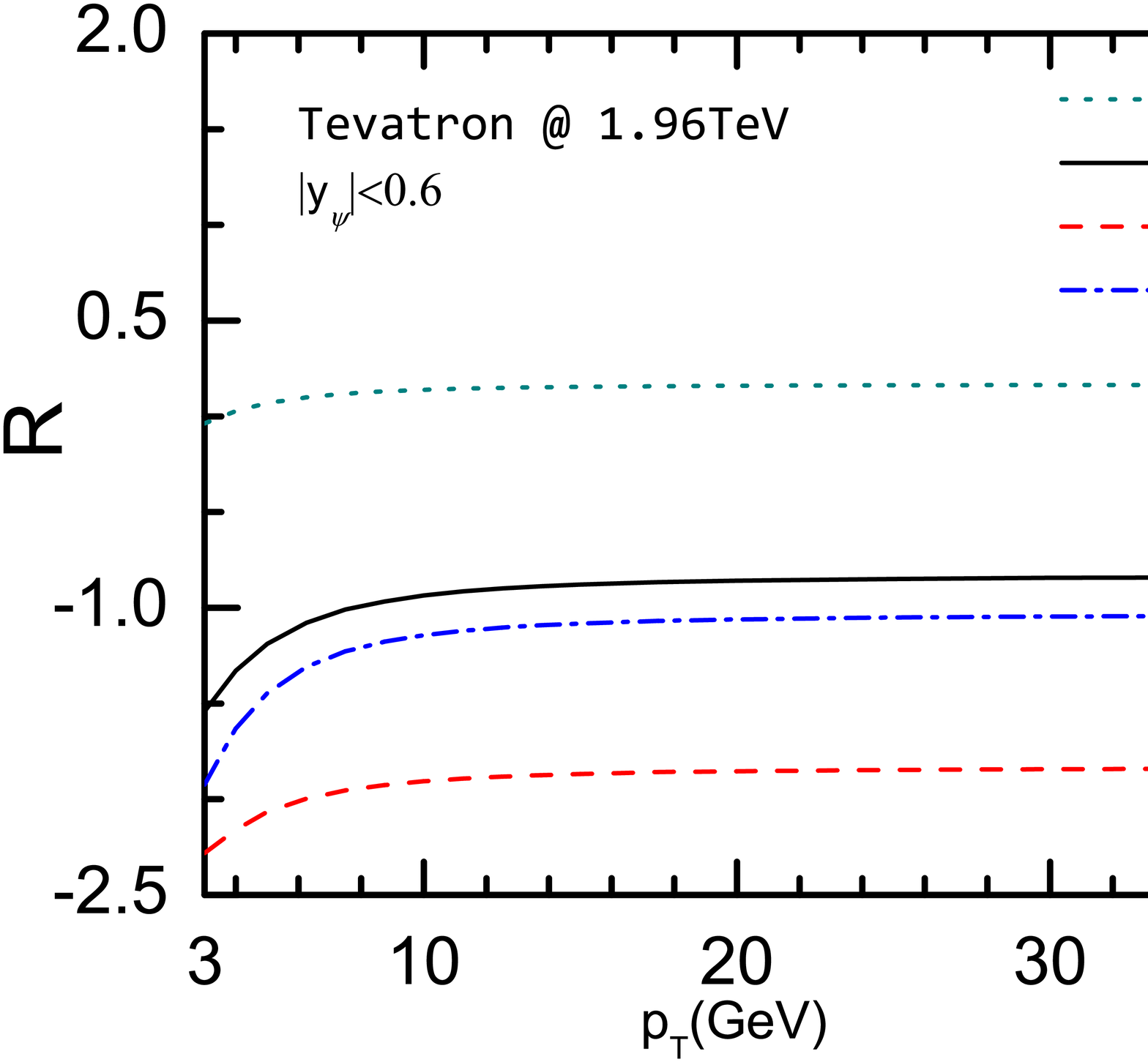}\hspace{1.7cm}
\includegraphics[width=0.4\textwidth]{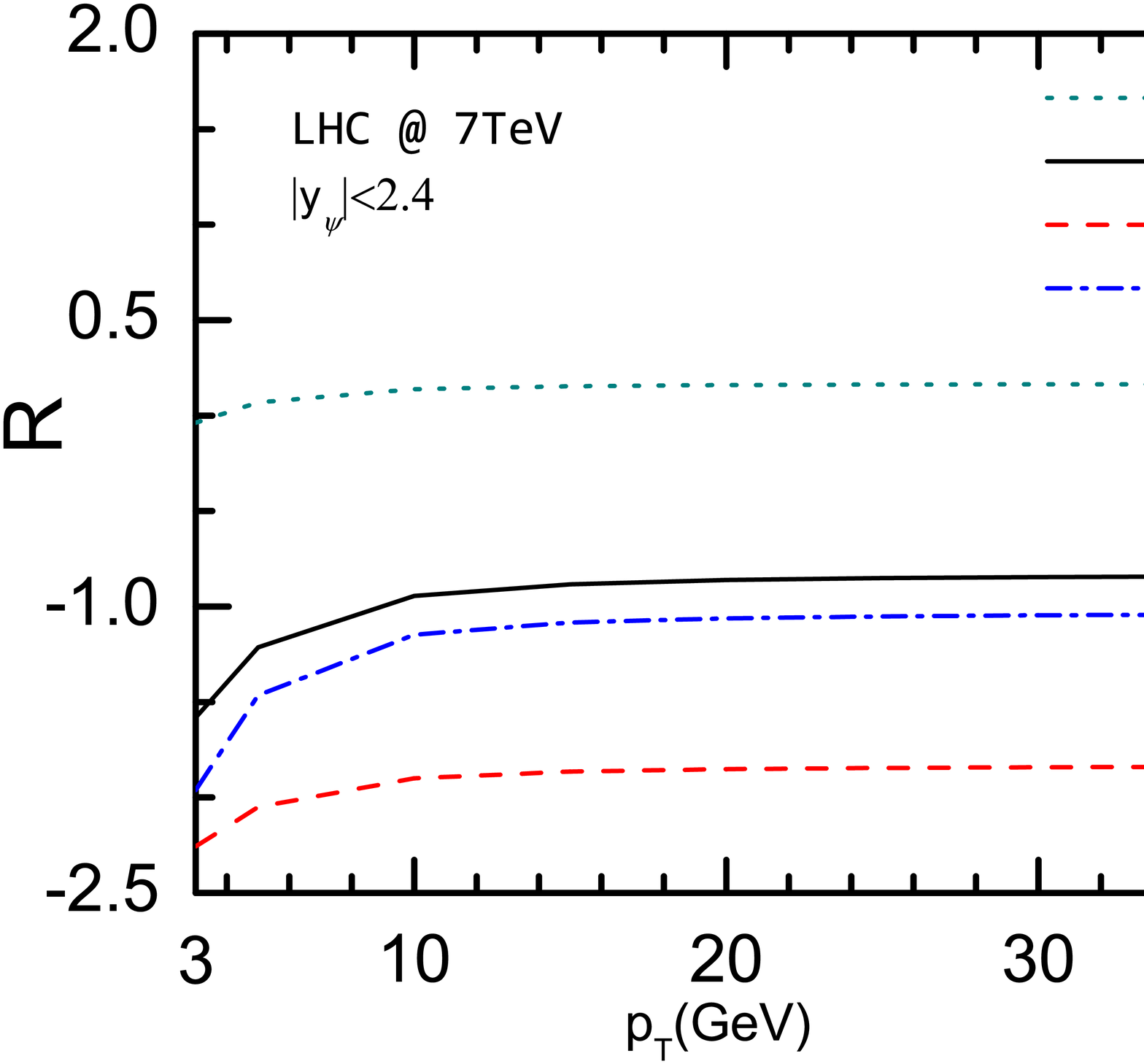}\\
\hspace{-1.4cm}\includegraphics[width=0.4\textwidth]{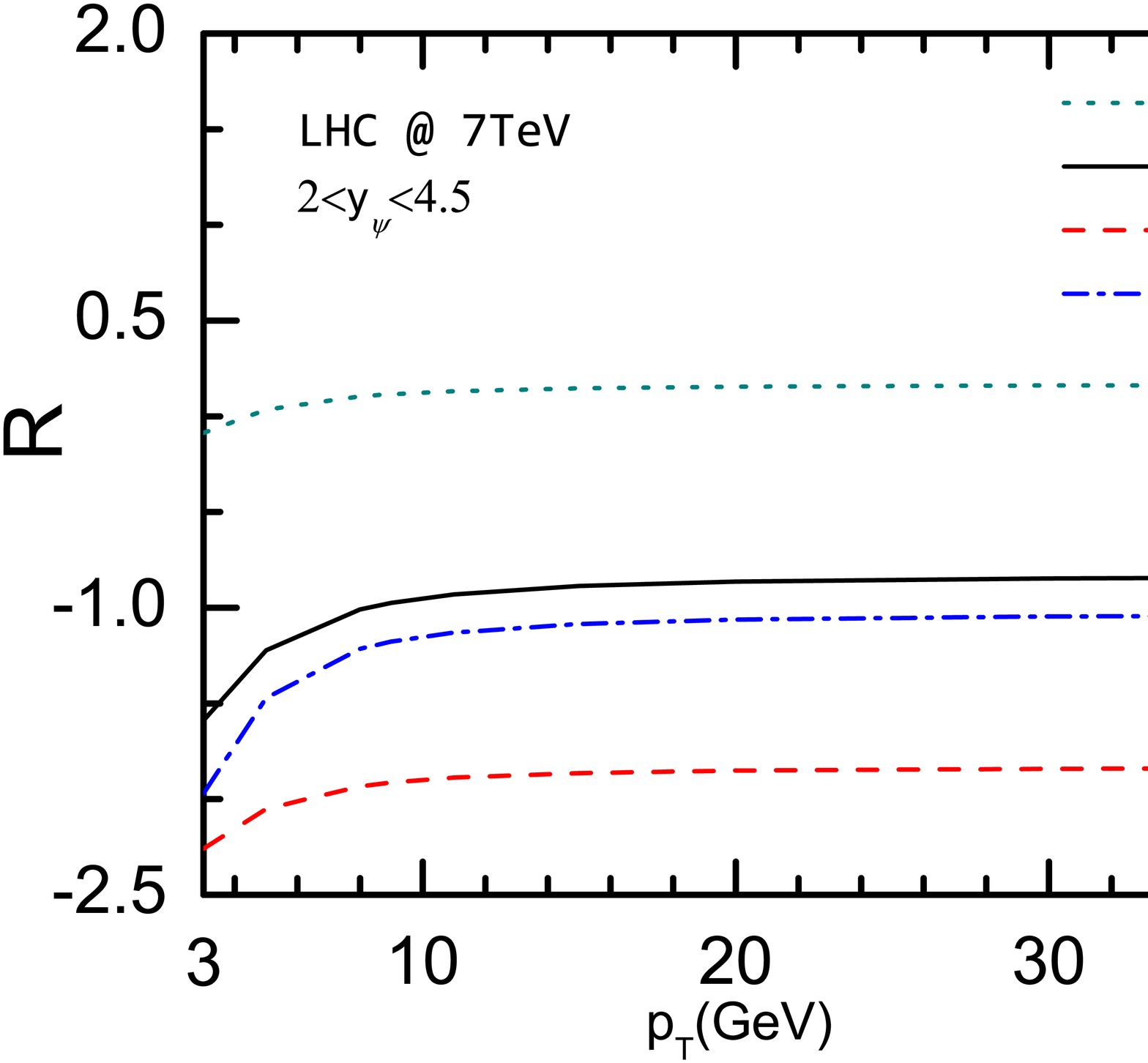}\hspace{1.7cm}
\includegraphics[width=0.4\textwidth]{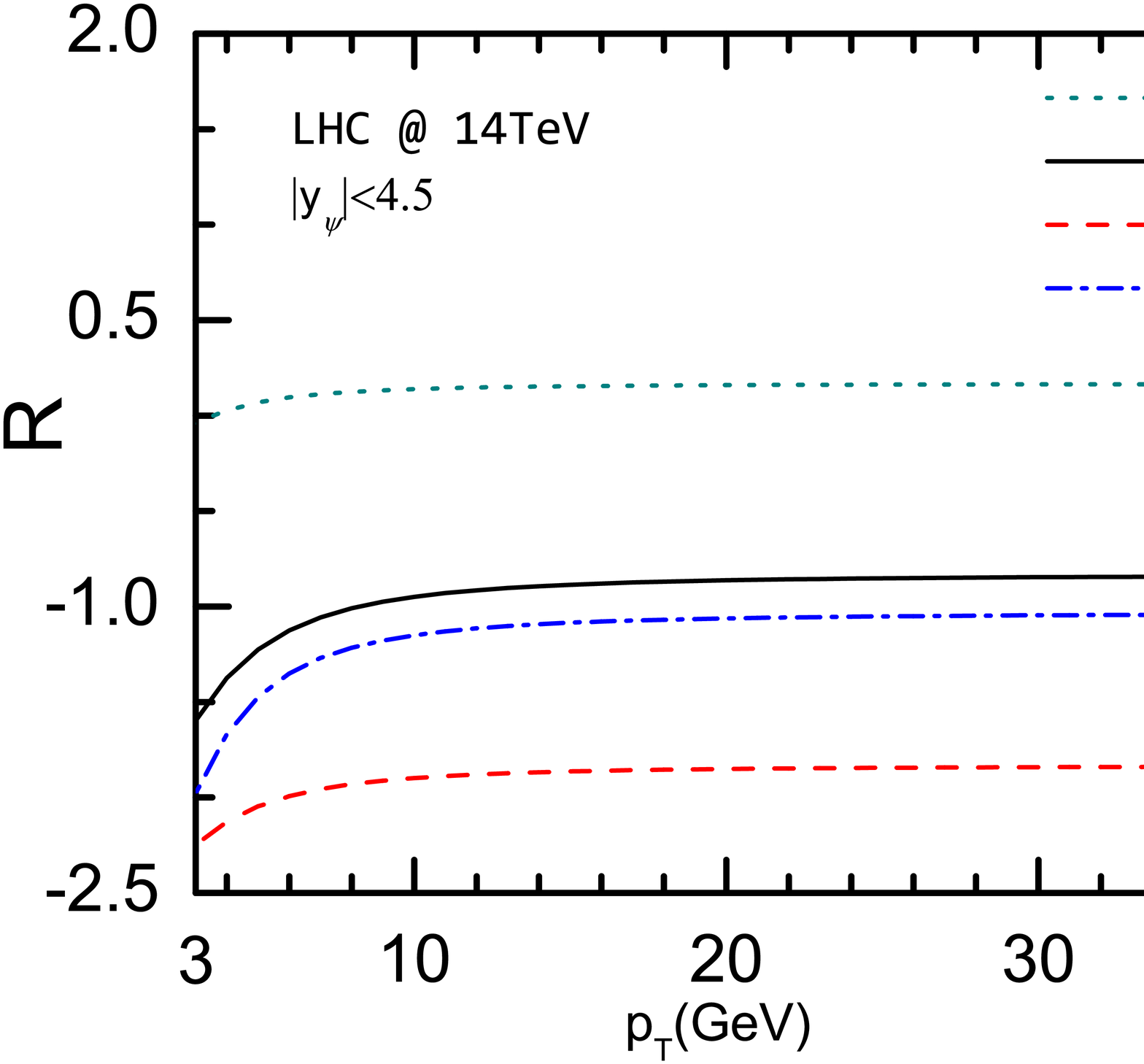}
\end{center}
\caption{\label{fig:kforsdc}The ratios of the short distance coefficient between LO $F$ and its
relativistic correction $G$ at the Tevatron with $\sqrt{s}=1.96~TeV$ and at
the LHC with $\sqrt{s}=7~TeV$ or $\sqrt{s}=14~TeV$.}
\end{figure}

We adopt the gluon distribution function CTEQ6 PDFs\cite{Pumplin:2002vw}. And the charm quark is set as $m_c=1.5~GeV$.
The ratios of the short distance coefficient between LO $F$ and
its
relativistic correction $G$ at the Tevatron with $\sqrt{s}=1.96~TeV$ and at
the LHC with $\sqrt{s}=7~TeV$ or $\sqrt{s}=14~TeV$ are presented in Fig.\ref{fig:kforsdc}. The ratios of $R[n]=G[n]/F[n]$ at the Tevatron
and at the LHC are very close at large $p_T$. In the large $p_T$ limit,
\begin{eqnarray}
-\frac{M^2}{u}\sim- \frac{M^2}{t}<\frac{M^2}{p_T^2}\sim0,
\end{eqnarray}
where $M$ is the $J/\psi$ mass. Then we can expand the short distance coefficients with $M$. The ratios of first order in the expansion
are
\begin{eqnarray}
R({}^3S_{1}^{[1]})\Big|_{p_T\gg M}=\frac{G({}^3S_1^{[1]})}{F({}^3S_1^{[1]})}\Big|_{p_T\gg M}&\sim&\frac{1}{6}\nonumber \\
R({}^1S_{0}^{[8]})\Big|_{p_T\gg M}=\frac{G({}^1S_0^{[8]})}{F({}^1S_0^{[8]})}\Big|_{p_T\gg M}&\sim&-\frac{5}{6}\nonumber \\
R({}^3S_{1}^{[8]})\Big|_{p_T\gg M}=\frac{G({}^3S_1^{[8]})}{F({}^3S_1^{[8]})}\Big|_{p_T\gg M}&\sim&-\frac{11}{6}\nonumber \\
R({}^3P_{0}^{[8]})\Big|_{p_T\gg M}=\frac{G({}^3P_0^{[8]})}{F({}^3P_0^{[8]})}\Big|_{p_T\gg M}&\sim&-\frac{31}{30}
\end{eqnarray}
These asymptotic behaviors of the ratios to each state are same for all the partonic subprocesses of $gg$, $gq(\bar{q})$ and $qq$.
It is consistent with the curves in Fig.1. The ratio $R({}^3S_1^{[1]})$ is consistent with Ref.\cite{Fan:2009zq}, and the ratio $R({}^3S_1^{[8]})$ is consistent with Ref.\cite{Bodwin:2003wh}.

 As discussed in Sec. II, the LDMEs
of relativistic correction are depressed by approximately 0.23 to LO.
If we fix LDMEs $\langle 0|\mathcal{O}|0\rangle$ and estimate $\langle 0|\mathcal{P}|0\rangle$ through the velocity scaling rule with adopting $v^2=0.23$,
then the LO cross sections of CO subprocesses are reduced by about a factor of $20\sim~40\%$ at large $p_T$ at both Tevatron and LHC.
In the CS case, the LO cross sections are enhanced by approximately $4\%$ by the NLO relativistic corrections. \footnote{In Ref.\cite{Fan:2009zq}, the ratio of the  CS cross sections enhanced by NLO relativistic corrections is about $1\%$ . The difference comes from adopting the different LDMEs:
\begin{equation}\label{matrixenhanced}
\langle
0|\mathcal{O}^{J/\psi}({}^3S_1^{[1]})|0\rangle=1.64~GeV^3,~~~~~~~\langle
0|\mathcal{P}^{J/\psi}({}^3S_1^{[1]})|0\rangle=0.320~GeV^5.
\end{equation}
Then
\begin{equation}\label{matrix}
\langle
0|\mathcal{P}^{J/\psi}({}^3S_1^{[1]})|0\rangle/\langle
0|\mathcal{O}^{J/\psi}({}^3S_1^{[1]})|0\rangle/m_c^2=0.087,
\end{equation}
which is much smaller than $v^2 \approx 0.23$.}

\begin{figure}[h]
\begin{center}
\hspace{-1.4cm}\includegraphics[width=0.4\textwidth]{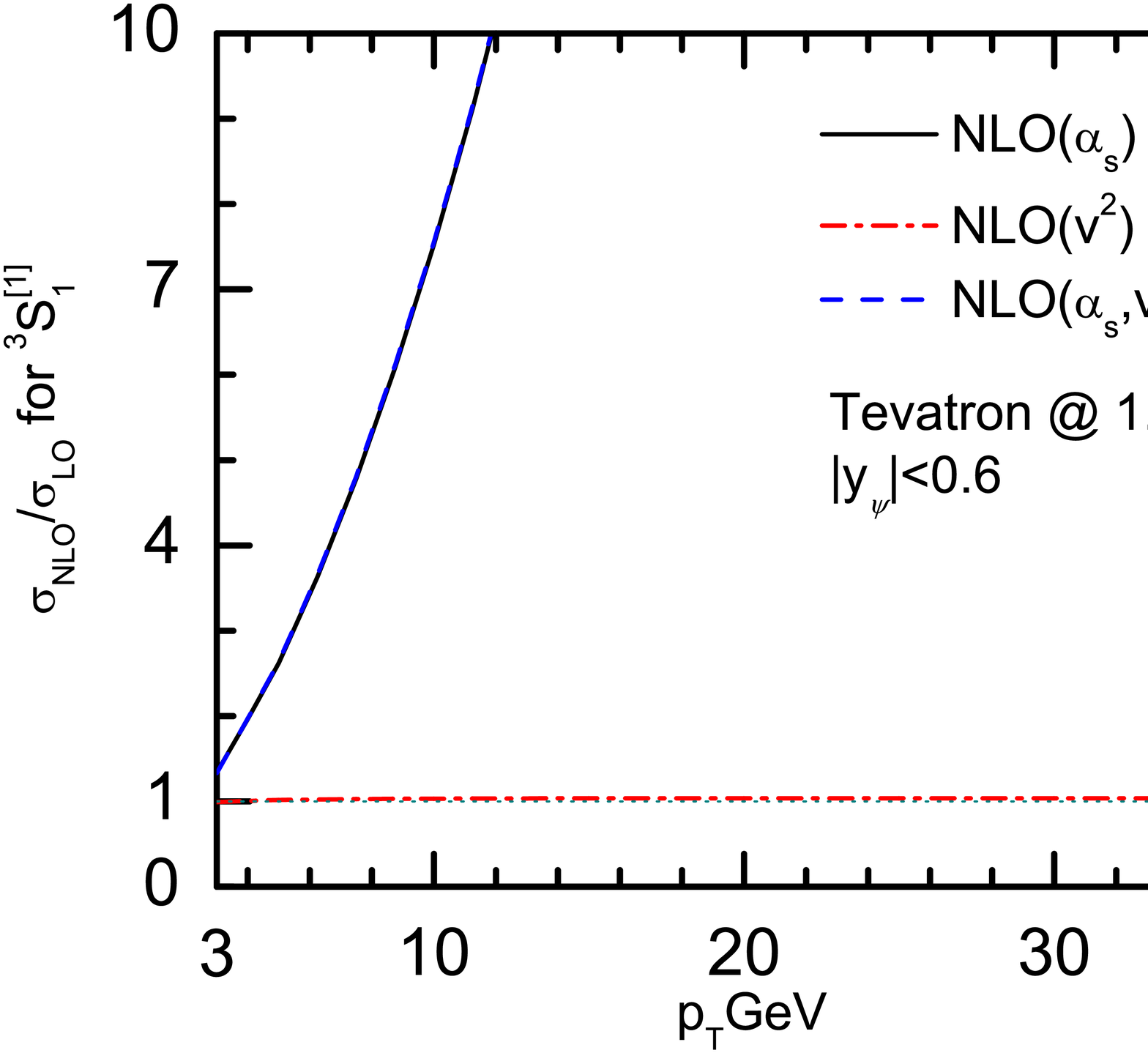}\hspace{1.7cm}
\includegraphics[width=0.4\textwidth]{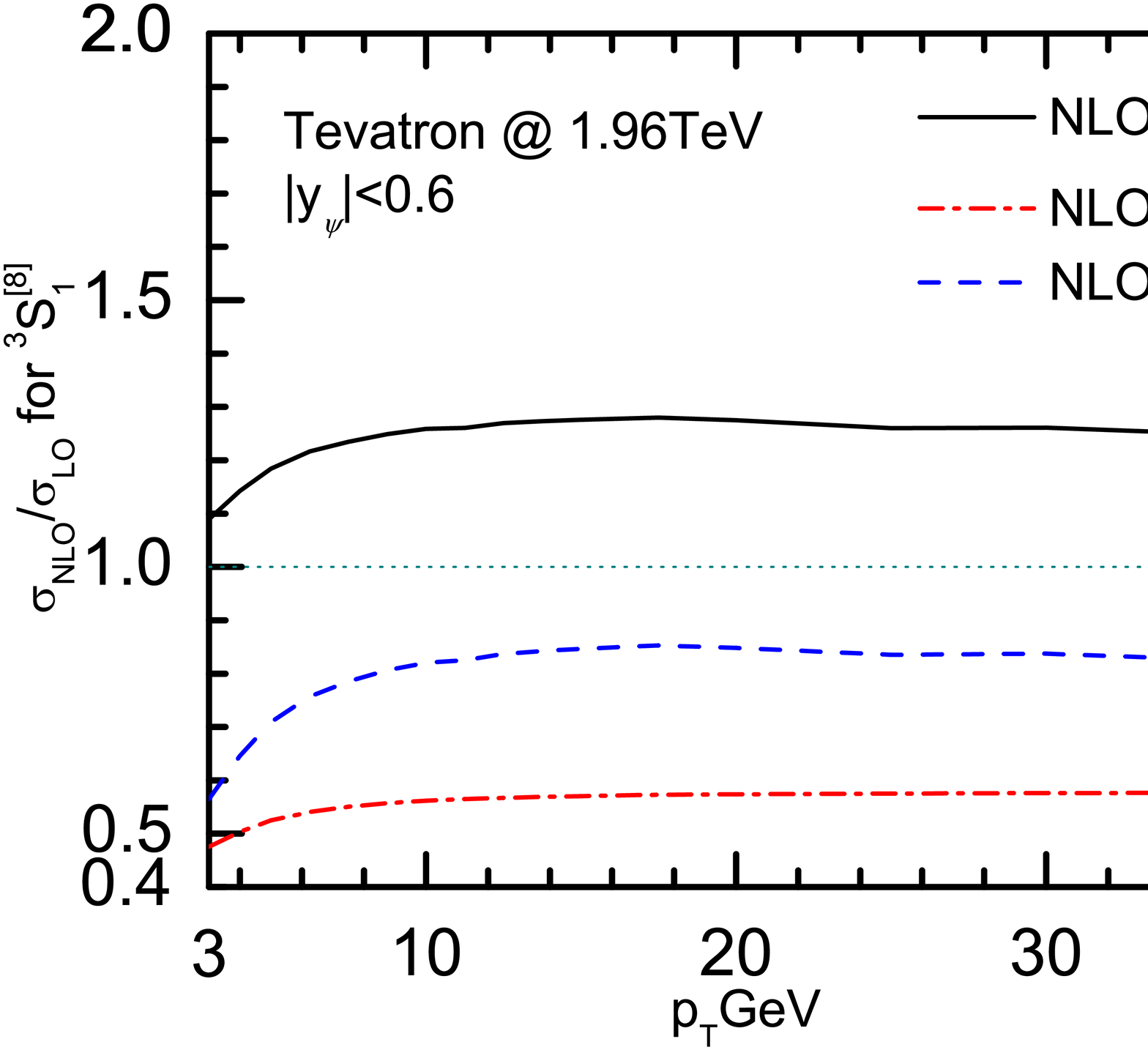}\\
\hspace{-1.4cm}\includegraphics[width=0.4\textwidth]{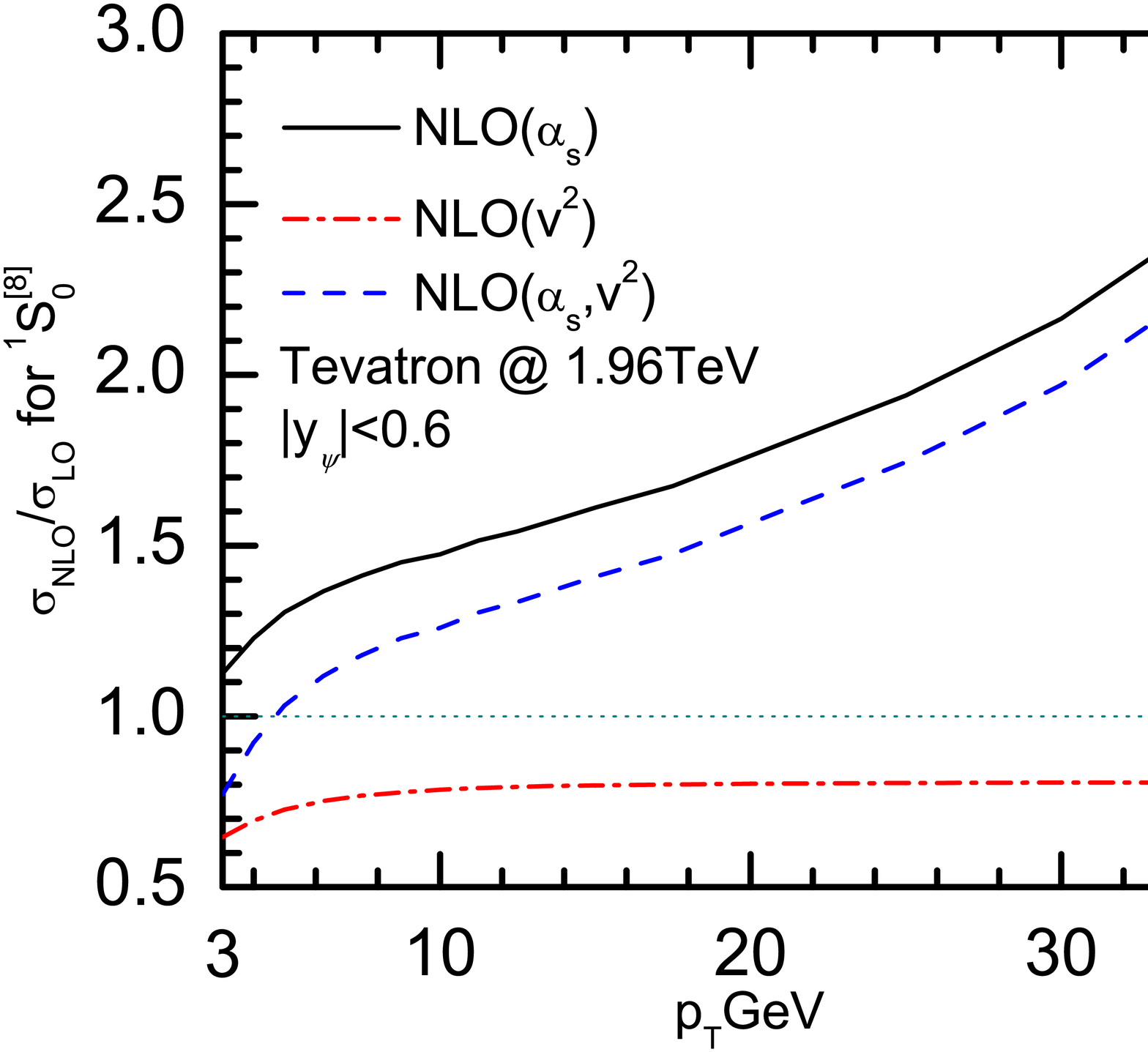}\hspace{1.7cm}
\includegraphics[width=0.4\textwidth]{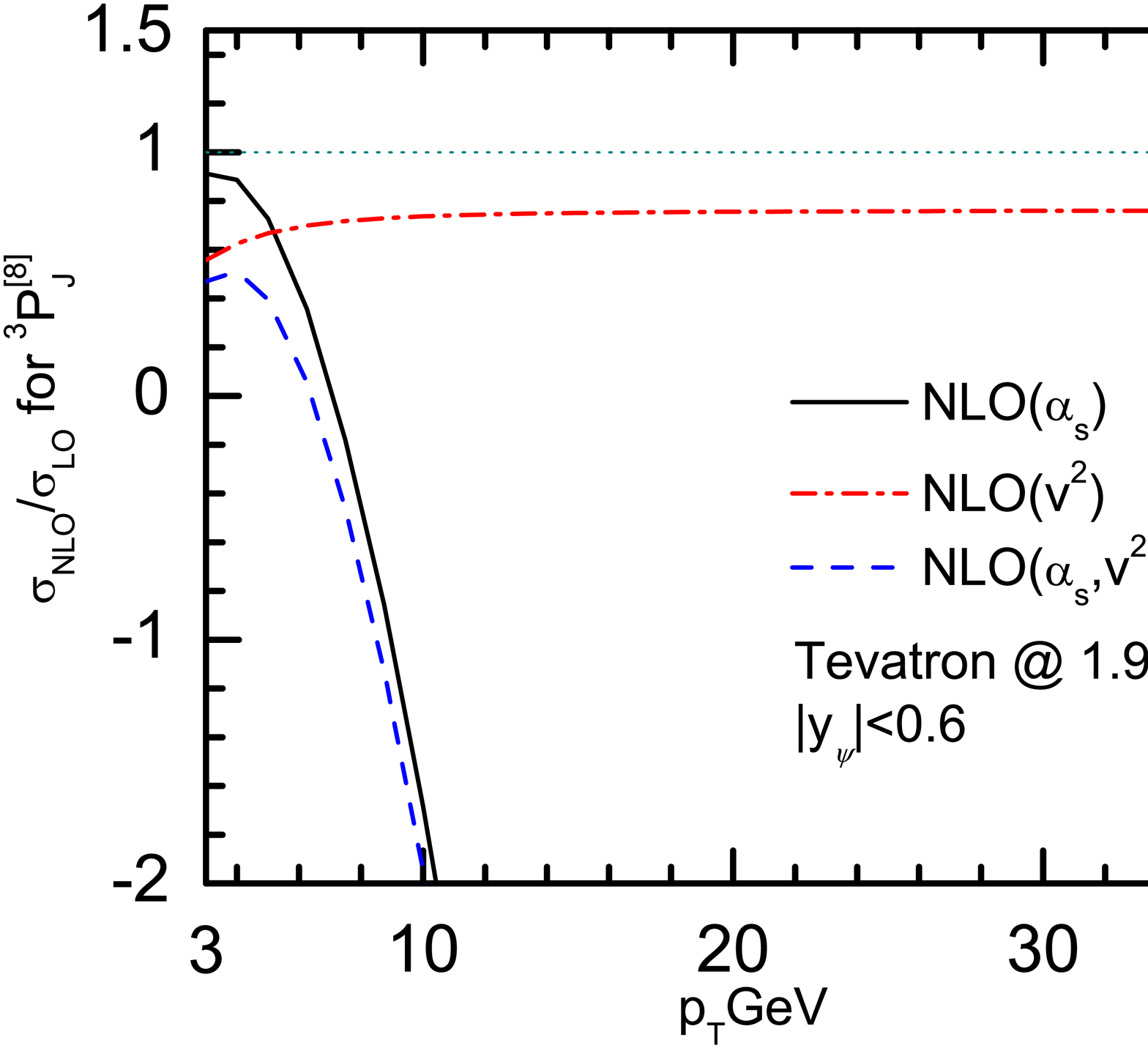}
\end{center}
\caption{\label{fig:kKTevatron}Ratios of NLO $\mathcal{O}(v^2)$, $\mathcal{O}(\alpha_s)$, and $\mathcal{O}(\alpha_s,v^2)$ to
LO cross sections  of $J/\psi$ production at Tevatron. Here $v^2=0.23$, and QCD corrections are taken form Refs.\cite{Ma:2010jj,Chao:2012iv}.}
\end{figure}

The QCD corrections of both CO and CS states had been calculated in \cite{Ma:2010jj,Chao:2012iv,Butenschoen:2011yh}. Ratios of NLO $\mathcal{O}(v^2)$, $\mathcal{O}(\alpha_s)$, and $\mathcal{O}(\alpha_s,v^2)$ to
LO cross sections  of $J/\psi$ production at Tevatron are presented in Fig.\ref{fig:kKTevatron}. Here $v^2=0.23$, and QCD corrections are taken
from
Refs.\cite{Ma:2010jj,Chao:2012iv,Butenschoen:2011yh}. The $K$ factor of NLO QCD corrections is very large for ~${}^3P_0^{[8]}$ ~and ~${}^3S_1^{[1]}$ at large $p_T$, and it is about $1.3$ for ~${}^3S_1^{[8]}$ and $1.5$ for ${}^1S_0^{[8]}$.

The ratio of ${}^3S_{1}^{[8]}$ is approximately $-11/6$. In the large $p_T$ limit, the dominate contribution of this subprocess is $g^*\to c\bar c ({}^3S_{1}^{[8]})$. The propagator of virtual gluon $g^*$ is proportional to $1/E_q^2$. This term offers a factor of $-2$ to the ratio $R({}^3S_{1}^{[8]})$. And the factor of $-2$ at large $p_T$ is same for the polarization of ${}^3S_{1}^{[8]}$ states.  At the same time, the ${}^1S_{0}^{[8]}$
state
is a scalar state and
contributes
to unpolarized production of $J/\psi$, and the $K$ factor of NLO QCD corrections is much larger than relativistic corrections for ~${}^3P_0^{[8]}$ ~and ~${}^3S_1^{[1]}$ at large $p_T$. So the $J/\psi$ polarization at large $p_T$ is insensitive to the relativistic corrections.

If we fit the differential cross
section of prompt $J/\psi$  production at $p_t> 7~ GeV$ at the
Tevatron \cite{arXiv:0704.0638} to NLO in $\alpha_s$ and $v^2$\cite{Ma:2010jj}, we can get CO LDMEs but with large errors for $^{3}S_1^{[8]}$ and $^{3}P_J^{[8]}$ states. In Ref.\cite{Ma:2010jj}, they considered two combined LDMEs to fit the data:
\begin{eqnarray}
&&M_{0,r_0}^{J/\psi}=<0|\mathcal{O}^{J/\psi}({}^1S_0^{[8]})|0>+\frac{r_0}{m_c^2}<0|\mathcal{O}^{J/\psi}({}^3P_0^{[8]})|0>,\nonumber\\
&&M_{1,r_1}^{J/\psi}=<0|\mathcal{O}^{J/\psi}({}^3S_1^{[8]})|0>+\frac{r_1}{m_c^2}<0|\mathcal{O}^{J/\psi}({}^3P_0^{[8]})|0>.
\end{eqnarray}
Here $r_0, r_1$ determined from the short distance coefficient decomposition holding within a small error
\begin{eqnarray}
d\hat{\sigma}[^{3}P_J^{[8]}]=r_0d\hat{\sigma}[^{1}S_0^{[8]}]+r_1\hat{\sigma}[^{3}S_1^{[8]}].
\end{eqnarray}
In Ref.\cite{Ma:2010jj}, they found $r_0=3.9$ and $r_1=-0.56$ using the NLO$(\alpha_s)$ results. When considering relativistic corrections as well as NLO$(\alpha_s)$ data we find $r_0=3.64$ and $r_1=-0.84$. Then we can fit CDF $J/\psi$ prompt production data to determine these two LDMEs as Fig. 3 showns. (Here, we do not consider the effect of the feed-down cross section form $\chi_{cJ}$ and ${\psi}\prime$ to the fit):
\begin{eqnarray}
&&M_{0,3.64}^{J/\psi}=(11.0\pm0.3) \times 10^{-2}GeV^3,\nonumber\\
&&M_{1,-0.84}^{J/\psi}=(0.16\pm0.02) \times 10^{-2}GeV^3,
\end{eqnarray}
comparing with fitting results only considering NLO$(\alpha_s)$ data
\begin{eqnarray}
&&M_{0,3.9}^{J/\psi}=(9.0\pm0.3) \times 10^{-2}GeV^3,\nonumber\\
&&M_{1,-0.56}^{J/\psi}=(0.13\pm0.02) \times 10^{-2}GeV^3.
\end{eqnarray}
About $20\%$ difference is shown for either LDMEs between the two sets.
\begin{figure}[h]
\begin{center}
\includegraphics[width=0.4\textwidth]{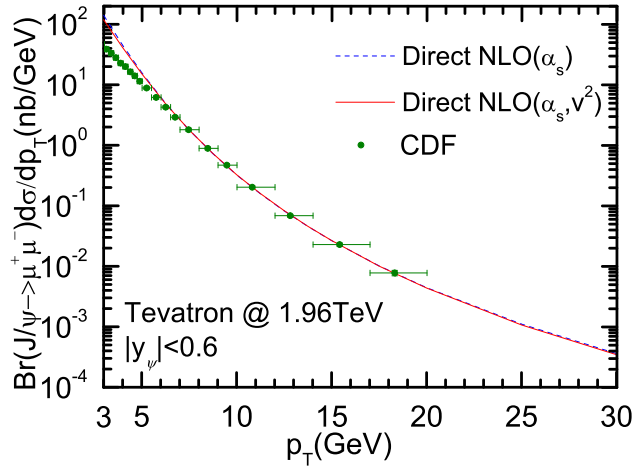}
\end{center}
\caption{\label{fig:cs_tevtron} Transverse momentum distribution of prompt $J/\psi$ production at Tevatron. By fitting the CDF experimental data we obtained the two sets of combined LDMEs $M_{0,r0}^{J/\psi}$ and $M_{1,r1}^{J/\psi}$ using the results of NLO$(\alpha_s)$ and NLO$(\alpha_s,v^2)$ short distance coefficients, respectively.}
\end{figure}
Complete NLO$(\alpha_s)$ calculations show the LDMEs fitting the Tevatron data agree with all the LHC data. However, it does not agree well at the small $p_T$ region \cite{Ma:2010jj}. The $K$ factor curves in Fig. 2 imply that relativistic corrections suppress the trend of the $K$ factors of NLO$(\alpha_s)$ mainly at small $p_T$ region.
To investigate the effect of new fitting LDMEs to the total cross section at hadron colliders, especially at small $p_T$ region, we compare the cross sections of NLO$(\alpha_s)$ and NLO$(\alpha_s,v^2)$ at the LHC using the corresponding set of LDMEs above, and the results are shown in Fig.4.
NLO$(\alpha_s,v^2)$ results suppressed by about $50\sim20\%$ along with $p_T$ increasing comparing with NLO$(\alpha_s)$ results.
But the calculations of relativistic correction of direct production fail to explain the tend of experimental data at the small $p_T$ region, and it is still an open problem. It is expected to solve the problem by two ways. First, contribution from the feed-down of high excited charmonia production process as $p+p(\bar{p}){\rightarrow}{\chi_{cJ}}+X$ and $p+p(\bar{p}){\rightarrow}{\psi\prime}+X$ may account for $30\%$ to prompt $J/\psi$ production. In this case, the calculations of relativistic correction to feed-down parts are necessary. Second, recently, the calculation method of resummation of relativistic correction had been presented by Bodwin, Lee and Yu and applied to calculate the resummation of relativistic correction to
exclusive production $e^{+}e^{-}\to J/\psi \eta_c$ at $e^{
+}
e^{-}$ colliders that payed an important contribution to total cross section\cite{Bodwin:2007ga}. Wether contributions of resummation of relativistic correction may play an important role, further calculations are needed.
\begin{figure}[h]
\begin{center}
\hspace{-1.4cm}\includegraphics[width=0.4\textwidth]{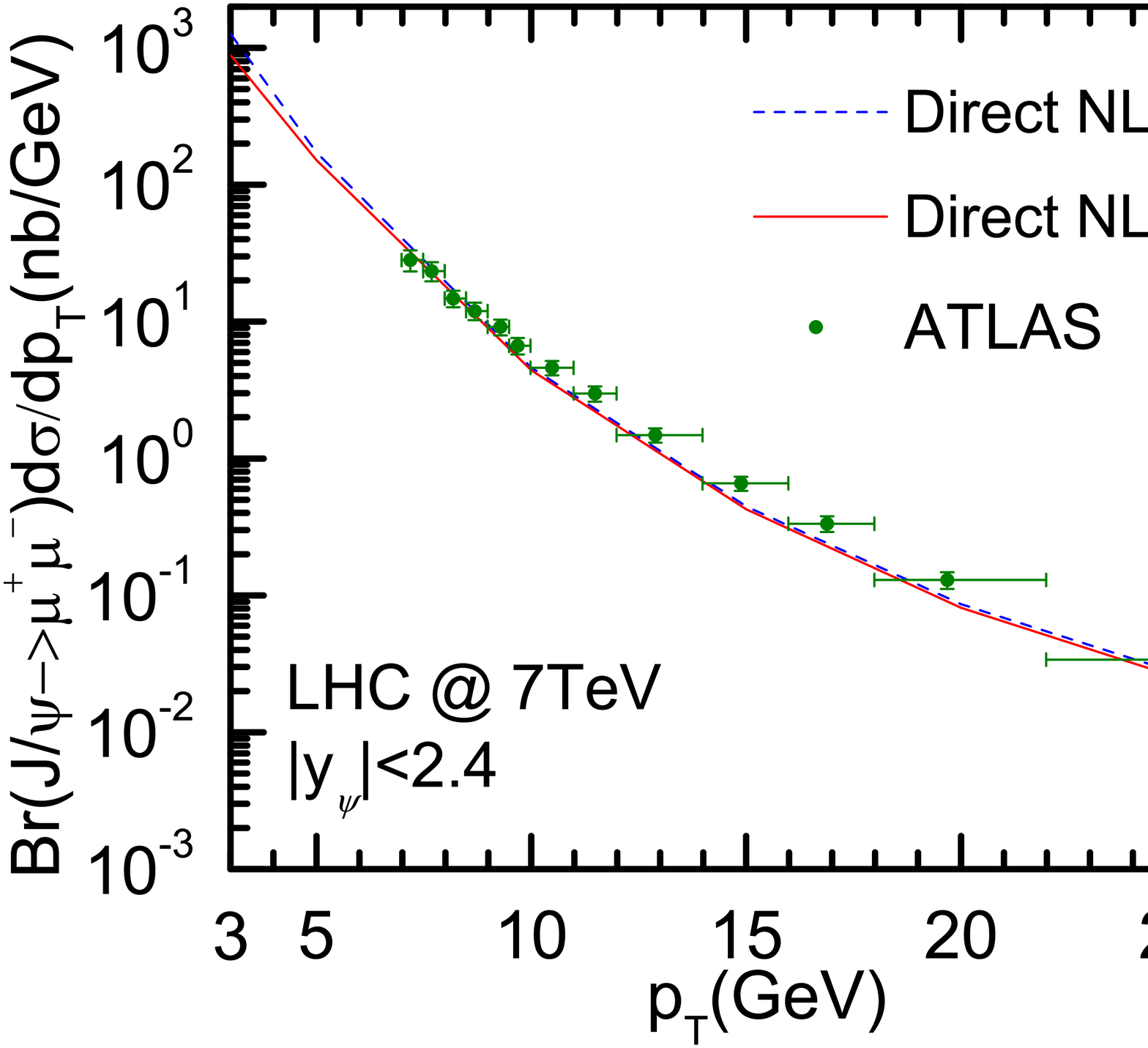}\hspace{1.7cm}
\includegraphics[width=0.4\textwidth]{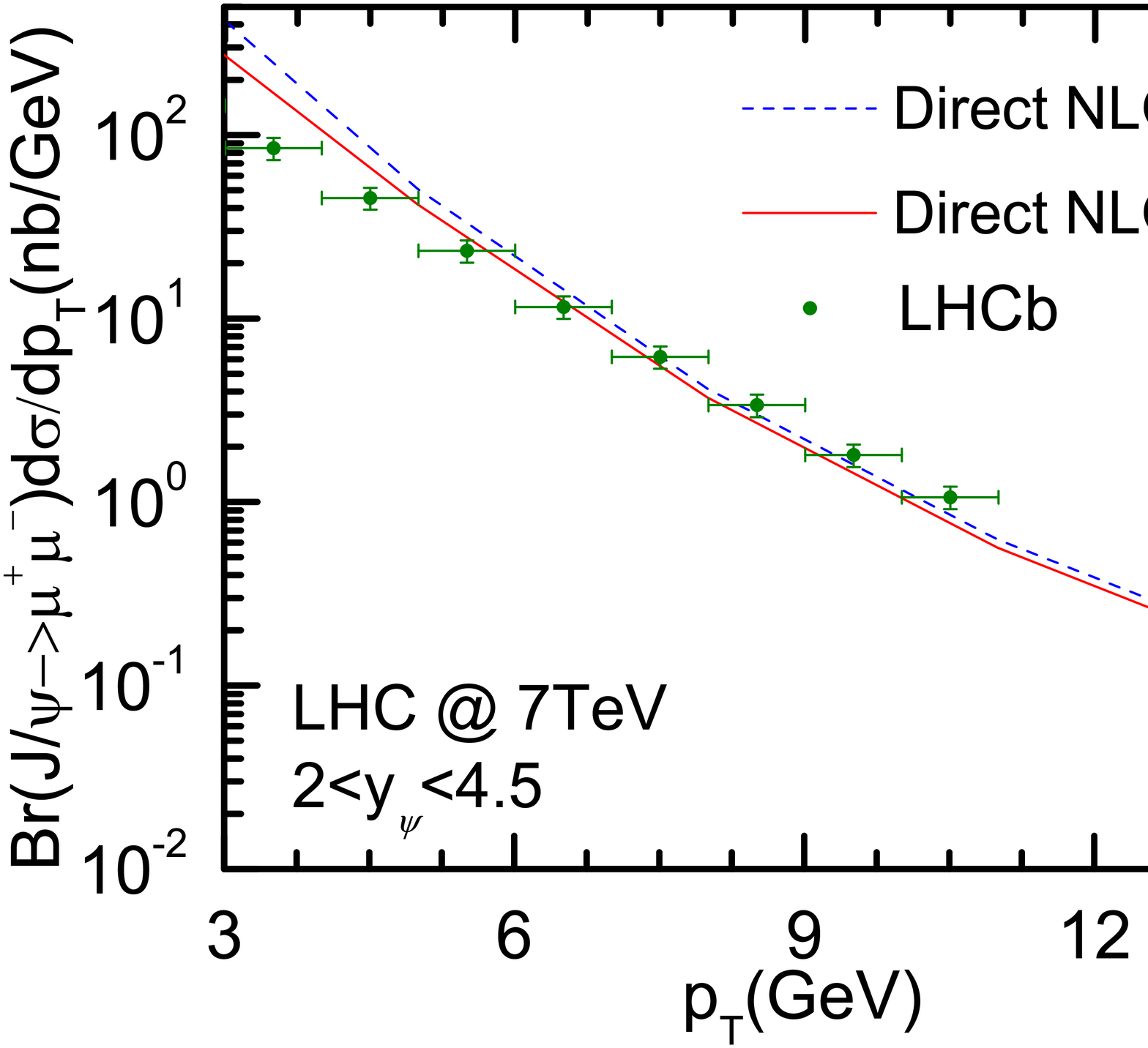}\\
\hspace{-1.4cm}\includegraphics[width=0.4\textwidth]{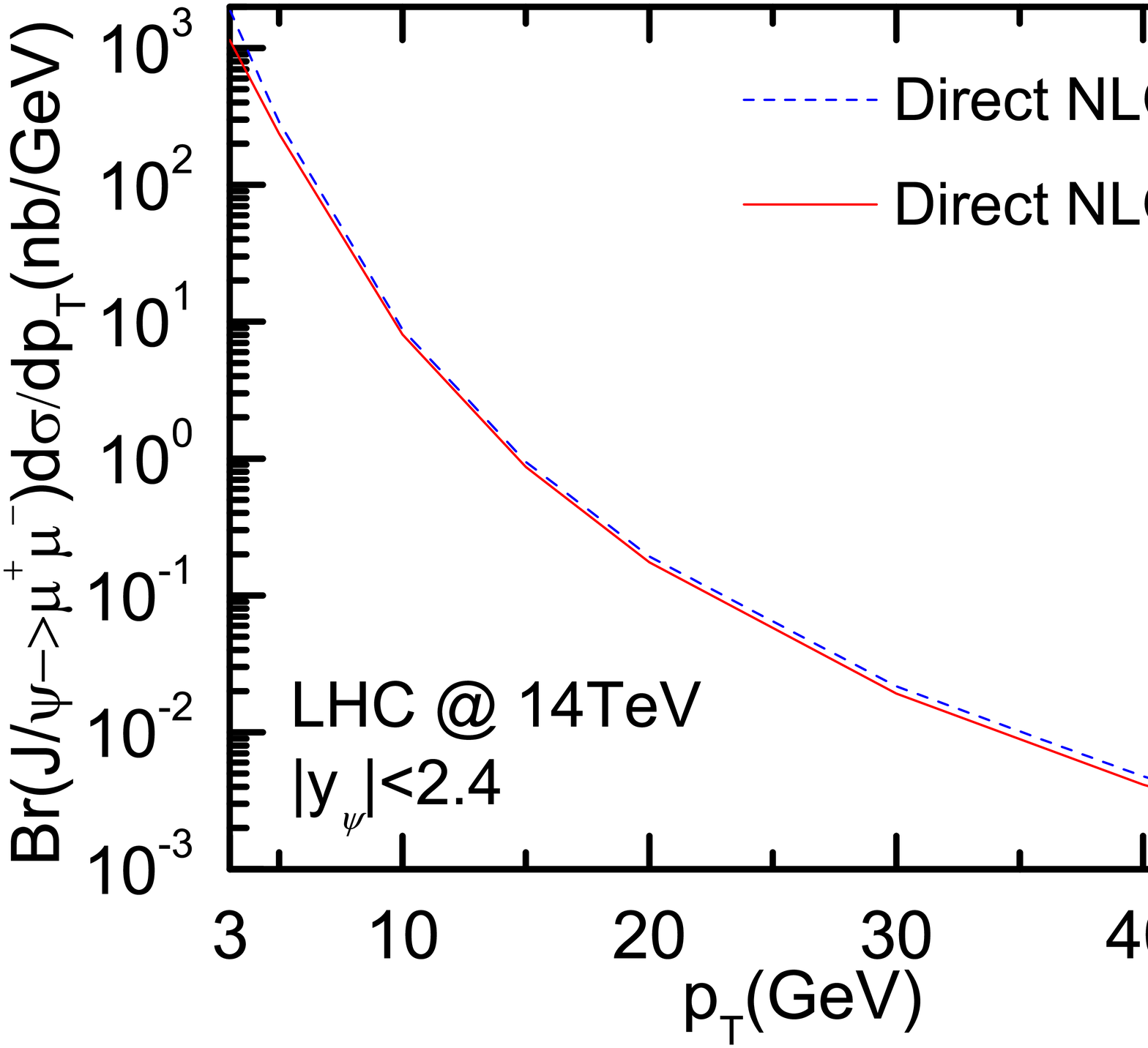}\hspace{1.7cm}
\includegraphics[width=0.4\textwidth]{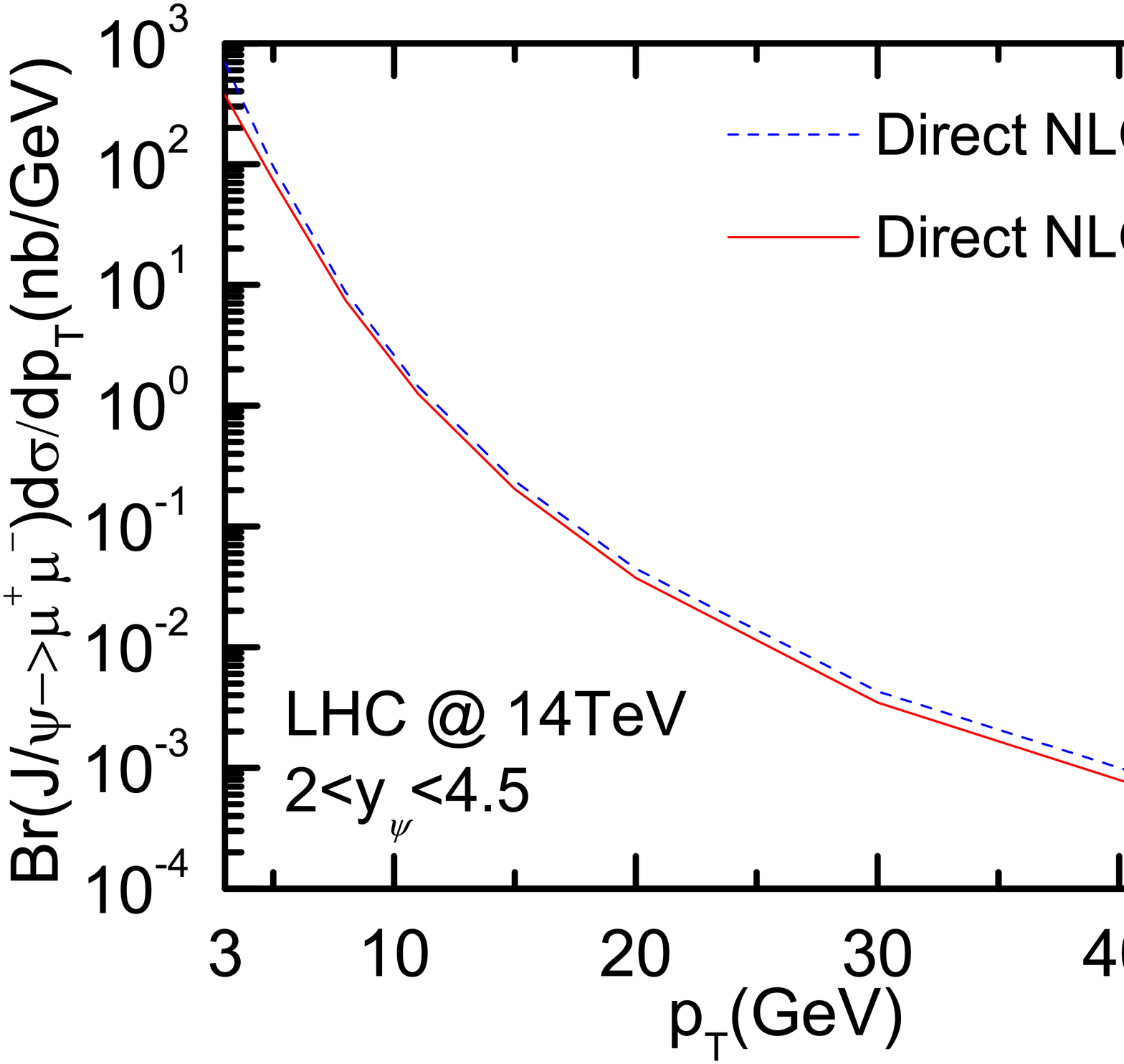}
\end{center}
\caption{\label{fig:cs lhc} Transverse momentum distribution of NLO$(\alpha_s)$ and NLO$(\alpha_s,v^2)$ to $J/\psi$ direct production. The LHC experimental data can be found in Refs.\cite{Aaij:2011jh,Aad:2011sp}.}
\end{figure}

\section{SUMMARY}
In summary, we calculate the relativistic correction terms
to CO states for $J/\psi$ production at the Tevatron and at
the LHC.
The short distance coefficient ratios of relativistic correction to LO
for CO states ${^1}S_{0}^{[8]}$, ${^3}S_{1}^{[8]}$ and ${^3}P_{J}^{[8]}$
at large $p_T$ are  approximately -5/6, -11/6, and -31/30, respectively,
and it is 1/6 for the color singlet-state  ${^3}S_{1}^{[1]}$.
If NLO long distance matrix elements are estimated through the velocity
scaling rule with adopting $v^2=0.23$,
the cross sections are reduced by about a factor of $20\sim40\%$ at large $p_T$ to LO results of CO states at both the Tevatron and the LHC.
Compared with the relativistic
corrections to the CS state, that LO cross sections are enhanced by a factor of $4\%$.  Thus the result may affect the
production of $J/\psi$ at hadronic colliders. Beacuse of the large results of QCD corrections at large $p_T$ especially to ${}^3P_J^{[8]}$ states, relativistic corrections are small, even ignored, along with $p_T$ increasing.
But relativistic corrections can also affect the total cross section with a considerable contribution.
We computed the unpolarized cross sections at the LHC with CO LDMEs extracted from the fit to $J/\psi$ direct production at the Tevatron, and the results of NLO$(\alpha_s,v^2)$  suppress that of NLO$(\alpha_s)$ by about $20\sim50\%$ at different $p_T$ regions.
These results indicate that relativistic corrections may play an important
role  in $J/\psi$ production at the Tevatron and LHC.

\begin{acknowledgments}
The authors would like to thank Professor K.T. Chao, Z.G. He,
Y.Q. Ma, H.S. Shao, and K. Wang for useful
discussion and the data of NLO QCD corrections. Y.J. Zhang also thanks J.P. Lansberg  for the discussion of polarization and B.Q. Li for the discussion of $v^2$.  This work was
supported by the National Natural Science
Foundation of China (Grants No.10805002, No.10875055, and
No.11075011), the Foundation for the Author of National
Excellent Doctoral Dissertation of China (Grants No. 2007B18 and No. 201020), the
Project of Knowledge Innovation Program (PKIP) of Chinese Academy of
Sciences, Grant No. KJCX2.YW.W10, and the Education Ministry of
Liaoning Province.

\end{acknowledgments}

\section{Appendix: Short Distance Coefficients}

The short distance coefficients of $^1S_0^{[8]}$
for $gg{\rightarrow}{J/\psi}g$ subprocess
were

\begin{eqnarray}
&& \hspace{-0.9cm}\frac{F_{gg}({}^1S_0^{[8]})}{m_c^2} =\frac{1}{16\pi
s^2}\frac{1}{64}\frac{1}{4}\frac{(4\pi\alpha_s)^3}{N_c^2-1} \times
\nonumber\\
&&640\Bigg[M^{12} \left(t^2+t u+u^2\right)-M^{10}
\left(4 t^3+7 t^2 u+7 t u^2+4 u^3\right)
\nonumber\\
&&+M^8 \left(8 t^4+21 t^3 u+27 t^2 u^2+21 t u^3+8 u^4\right)-M^6
\left(10 t^5+35 t^4 u+57 t^3 u^2+57 t^2 u^3+35
t u^4+10 u^5\right)
\nonumber\\
&&+M^4 \left(8 t^6+33 t^5 u+66 t^4 u^2+81 t^3 u^3+
66 t^2 u^4+33 t u^5+8 u^6\right)
\nonumber\\
&&-M^2 \hspace{-0.1cm} \left(t^2+t u+u^2\right)^2
\hspace{-0.1cm}\left(4 t^3+9 t^2 u+9 t u^2+4 u^3\right)\hspace{-0.1cm}+\left(t^2+t u+u^2\right)^4
\Bigg]
\nonumber\\
&&\Bigg/\Bigg[M \left(M^2-t\right)^2
t \left(M^2-u\right)^2 \left(M^2-t-u\right) u (t+u)^2\Bigg],
\end{eqnarray}

\begin{eqnarray}
&& \hspace{-0.9cm}\frac{G_{gg}({}^1S_0^{[8]})}{m_c^4}=\frac{1}{16\pi
s^2}\frac{1}{64}\frac{1}{4}\frac{(4\pi\alpha_s)^3}{N_c^2-1}
\nonumber\\
&&1280 \Bigg[5 t u (t+u)
\left(t^2+t u+u^2\right)^4 + 12 M^{18} \left(t^2+t u+u^2\right)-5 M^{16} \left(11 t^3+20 t^2 u+20 t u^2+11 u^3\right)
\nonumber\\
&&+M^{14} \left(95 t^4+280 t^3 u+358 t^2 u^2+280 t u^3+95
u^4\right)
\nonumber\\
&&-3 M^{12} \left(16 t^5+95 t^4 u+175 t^3 u^2+175 t^2 u^3+95 t u^4+16 u^5\right)
\nonumber\\
&&-2 M^{10} \left(45 t^6+72 t^5 u+21 t^4 u^2-22 t^3 u^3+21
t^2 u^4+72 t u^5+45 u^6\right)
\nonumber\\
&&+M^8 \left(198 t^7+678 t^6 u+1141 t^5 u^2+1345 t^4 u^3+
1345 t^3 u^4+1141 t^2 u^5+678 t u^6+198 u^7\right)
\nonumber\\
&&-M^6 \hspace{-0.1cm} \left(\hspace{-0.1cm}  180
t^8  \hspace{-0.1cm}+  \hspace{-0.1cm} 756 t^7 u  \hspace{-0.1cm}+
  \hspace{-0.1cm} 1583 t^6 u^2
\hspace{-0.1cm}+\hspace{-0.1cm}  2224 t^5 u^3 \hspace{-0.1cm}
+\hspace{-0.1cm}2446 t^4 u^4+2224 t^3 u^5+1583 t^2 u^6+756 t u^7+180 u^8\right)
\nonumber\\
&&+M^4 (85 t^9+408 t^8 u+1000 t^7 u^2+1637
t^6 u^3+2028 t^5 u^4+2028 t^4 u^5+1637 t^3 u^6+1000 t^2 u^7
\nonumber\\
&&+408 t u^8+85 u^9)-M^2
\left(t^3+2 t^2 u+2 t u^2+u^3\right)^2
\left(17 t^4+30 t^3 u+30 t^2 u^2+30 t u^3+17 u^4\right)\Bigg]\Bigg/
\nonumber\\
&&\Bigg[3 M^3 \left(M^2-t\right)^3 t \left(M^2-u\right)^3 u (t+u)^3
\left(-M^2+t+u\right)\Bigg].
\end{eqnarray}

The short distance coefficients of $^3S_1^{[8]}$
for $gg{\rightarrow}{J/\psi}g$ subprocess
were

\begin{eqnarray}
&& \hspace{-0.9cm}\frac{F_{gg}({}^3S_1^{[8]})}{m_c^2}=\frac{1}{16\pi
s^2}\frac{1}{64}\frac{1}{4}\frac{(4\pi\alpha_s)^3}{N_c^2-1}\frac{1}{3}
\nonumber\\
&&256 \Bigg[27 \left(t^2+t u+u^2\right)^3 + 19 M^8 \left(t^2+t u+u^2\right)-M^6 \left(65 t^3+111 t^2 u+111 t u^2+65 u^3\right)
\nonumber\\
&&+M^4 \left(100
t^4+227 t^3 u+300 t^2 u^2+227 t u^3+100 u^4\right)
\nonumber\\
&&-27 M^2 \left(3 t^5+8 t^4 u+13 t^3 u^2+13 t^2 u^3+8 t u^4+3 u^5\right)
\Bigg]
\nonumber\\
&&\Bigg/\Bigg[3 M^3 \left(M^2-t\right)^2
\left(M^2-u\right)^2 (t+u)^2\Bigg],
\end{eqnarray}
\begin{eqnarray}
&& \hspace{-0.9cm}\frac{G_{gg}({}^3S_1^{[8]})}{m_c^4}=\frac{1}{16\pi
s^2}\frac{1}{64}\frac{1}{4}\frac{(4\pi\alpha_s)^3}{N_c^2-1}\frac{1}{3}
\nonumber\\
&&(-512) \Bigg[M^{14} \left(87 t^2+22 t u+87 u^2\right)+M^{12}
\left(-14 t^3+335 t^2 u+335 t u^2-14
u^3\right)
\nonumber\\
&&-2 M^{10} \left(399 t^4+1612 t^3 u+2020 t^2 u^2+1612 t u^3+399 u^4\right)
\nonumber\\
&&+M^8 \left(2100 t^5+8976 t^4 u+14497 t^3 u^2+14497 t^2 u^3+8976
t u^4+2100 u^5\right)
\nonumber\\
&&-M^6 \left(2590 t^6+12096 t^5 u+23855 t^4 u^2+29314 t^3 u^3
+23855 t^2 u^4+12096 t u^5+2590 u^6\right)
\nonumber\\
&&+M^4 \left(1620 t^7+8498
t^6 u+19905 t^5 u^2+29152 t^4 u^3+29152 t^3 u^4+19905 t^2 u^5
+8498 t u^6+1620 u^7\right)
\nonumber\\
&&-27 M^2 \left(15 t^8+104 t^7 u+295 t^6 u^2+510 t^5 u^3+612
t^4 u^4+510 t^3 u^5+295 t^2 u^6+104 t u^7+15 u^8\right)
\nonumber\\
&&+297 t u (t+u) \left(t^2+t u+u^2\right)^3\Bigg]\Bigg/
\Bigg[9 M^5 \left(M^2-t\right)^3 \left(M^2-u\right)^3 (t+u)^3\Bigg].
\end{eqnarray}

The short distance coefficients of $^3P_J^{[8]}$
for $gg{\rightarrow}{J/\psi}g$ subprocess
were
\begin{eqnarray}
&& \hspace{-0.9cm}\frac{F_{gg}({}^3P_J^{[8]})}{m_c^4} =\frac{1}{16\pi
s^2}\frac{1}{64}\frac{1}{4}\frac{(4\pi\alpha_s)^3}{N_c^2-1} \times
\nonumber\\
&&2560 \bigg[7 M^{16} \left(t^3+2 t^2 u+2 t
   u^2+u^3\right)-M^{14} \left(35 t^4+99 t^3 u+120 t^2
   u^2+99 t u^3+35 u^4\right)
\nonumber\\
&&+M^{12} \left(84 t^5+296 t^4
   u+450 t^3 u^2+450 t^2 u^3+296 t u^4+84 u^5\right)
\nonumber\\
&&-3 M^{10} \left(42 t^6+171 t^5 u+304 t^4 u^2+362 t^3
   u^3+304 t^2 u^4+171 t u^5+42 u^6\right)
\nonumber\\
&&+M^8 \left(126
   t^7+577 t^6 u+1128 t^5 u^2+1513 t^4 u^3+1513 t^3
   u^4+1128 t^2 u^5+577 t u^6+126 u^7\right)
\nonumber\\
&&-M^6 \left(84
   t^8+432 t^7 u+905 t^6 u^2+1287 t^5 u^3+1436 t^4 u^4+1287
   t^3 u^5+905 t^2 u^6+432 t u^7+84 u^8\right)
\nonumber\\
&&+M^4 \left(35
   t^9+204 t^8 u+468 t^7 u^2+700 t^6 u^3+819 t^5 u^4+819
   t^4 u^5+700 t^3 u^6+468 t^2 u^7+204 t u^8+35
   u^9\right)
\nonumber\\
&&-M^2 \left(t^2+t u+u^2\right)^2 \left(7 t^6+36
   t^5 u+45 t^4 u^2+28 t^3 u^3+45 t^2 u^4+36 t u^5+7
   u^6\right)
\nonumber\\
&&+3 t u (t+u) \left(t^2+t u+u^2\right)^4\bigg]\bigg/\bigg[{M^3 t u \left(M^2-t\right)^3
   \left(M^2-u\right)^3 (t+u)^3 \left(-M^2+t+u\right)}\bigg].
\end{eqnarray}

\begin{eqnarray}
&& \hspace{-0.9cm}\frac{G_{gg}({}^3P_J^{[8]})}{m_c^6} =\frac{1}{16\pi
s^2}\frac{1}{64}\frac{1}{4}\frac{(4\pi\alpha_s)^3}{N_c^2-1} \times
\nonumber\\
&&(-1024) \bigg[140 M^{22} \left(t^3+2 t^2 u+2 t
   u^2+u^3\right)-M^{20} \left(725 t^4+2095 t^3 u+2596 t^2
   u^2+2095 t u^3+725 u^4\right)
\nonumber\\
&&+6 M^{18} \left(235 t^5+978
   t^4 u+1599 t^3 u^2+1599 t^2 u^3+978 t u^4+235
   u^5\right)
\nonumber\\
&&-M^{16} \left(705 t^6+6528 t^5 u+16050 t^4
   u^2+20350 t^3 u^3+16050 t^2 u^4+6528 t u^5+705
   u^6\right)
\nonumber\\
&&+M^{14} \left(-2190 t^7-3022 t^6 u+5603 t^5
   u^2+15689 t^4 u^3+15689 t^3 u^4+5603 t^2 u^5-3022 t
   u^6-2190 u^7\right)
\nonumber\\
&&+M^{12} (5400 t^8+19278 t^7
   u+25697 t^6 u^2+19598 t^5 u^3+14174 t^4 u^4
\nonumber\\
&&+19598 t^3
   u^5+25697 t^2 u^6+19278 t u^7+5400 u^8)
\nonumber\\
&&-M^{10}
   (6110 t^9+28087 t^8 u+52760 t^7 u^2+62879 t^6
   u^3+60308 t^5 u^4+60308 t^4 u^5
\nonumber\\
&&+62879 t^3 u^6+52760 t^2
   u^7+28087 t u^8+6110 u^9)
\nonumber\\
&&+M^8 (4055
   t^{10}+22235 t^9 u+50834 t^8 u^2+74420 t^7 u^3+83867 t^6
   u^4+84706 t^5 u^5
\nonumber\\
&&+83867 t^4 u^6+74420 t^3 u^7+50834 t^2
   u^8+22235 t u^9+4055 u^{10})
\nonumber\\
&&-M^6 (1530
   t^{11}+10029 t^{10} u+27765 t^9 u^2+49691 t^8 u^3+67682
   t^7 u^4+76683 t^6 u^5
\nonumber\\
&&+76683 t^5 u^6+67682 t^4 u^7+49691
   t^3 u^8+27765 t^2 u^9+10029 t u^{10}+1530
   u^{11})
\nonumber\\
&&+M^4 (255 t^{12}+2250 t^{11} u+8158
   t^{10} u^2+18865 t^9 u^3+32387 t^8 u^4+43880 t^7
   u^5+48446 t^6 u^6
\nonumber\\
&&+43880 t^5 u^7+32387 t^4 u^8+18865 t^3
   u^9+8158 t^2 u^{10}+2250 t u^{11}+255 u^{12})
\nonumber\\
&&-M^2
   t u \left(t^2+t u+u^2\right)^2 (150 t^7+726 t^6
   u+1575 t^5 u^2+2117 t^4 u^3+2117 t^3 u^4+1575 t^2
   u^5+726 t u^6
\nonumber\\
&&+150 u^7)+31 t^2 u^2 (t+u)^2
   \left(t^2+t u+u^2\right)^4\bigg]
\nonumber\\
&&\bigg/\bigg[{M^5 t u
   \left(M^2-t\right)^4 \left(M^2-u\right)^4 (t+u)^4
   \left(-M^2+t+u\right)}\bigg].
\end{eqnarray}

The short distance coefficients of $^1S_0^{[8]}$ for $q\bar{q}{\rightarrow}{J/\psi}g$ subprocess were
\begin{equation}
 \hspace{-0.9cm}\frac{F_{q\bar{q}}({}^1S_0^{[8]})}{m_c^2}=-\frac{1}{16\pi
s^2}\frac{1}{9}\frac{1}{4}\frac{(4\pi\alpha_s)^3}{N_c^2-1}\frac{160 \left(t^2+u^2\right)}{3 M (t+u)^2 \left(-M^2+t+u\right)}.
\end{equation}
\begin{eqnarray}
 &&\hspace{-0.9cm}\frac{G_{q\bar{q}}({}^1S_0^{[8]})}{m_c^4}=\frac{1}{16\pi
s^2}\frac{1}{9}\frac{1}{4}\frac{(4\pi\alpha_s)^3}{N_c^2-1}\frac{1600 \left(t^2+u^2\right)}{9 M^3 (t+u)^2
   \left(-M^2+t+u\right)}.
\end{eqnarray}
The short distance coefficients of $^3S_1^{[8]}$ for $q\bar{q}{\rightarrow}{J/\psi}g$ subprocess were
\begin{eqnarray}
 &&\hspace{-0.9cm}\frac{F_{q\bar{q}}({}^3S_1^{[8]})}{m_c^2}=-\frac{1}{16\pi
s^2}\frac{1}{9}\frac{1}{4}\frac{(4\pi\alpha_s)^3}{N_c^2-1}\frac{1}{3}\frac{64 \left(4 t^2-t u+4 u^2\right) \left(2 M^4-2 M^2
   (t+u)+t^2+u^2\right)}{3 M^3 t u (t+u)^2}.
\end{eqnarray}
\begin{eqnarray}
&& \hspace{-0.9cm}\frac{G_{q\bar{q}}({}^3S_1^{[8]})}{m_c^4}=\frac{1}{16\pi
s^2}\frac{1}{9}\frac{1}{4}\frac{(4\pi\alpha_s)^3}{N_c^2-1}\frac{1}{3}
\nonumber\\
&&(-128) \bigg[24 M^6 \left(4 t^2-t u+4 u^2\right)-14 M^4
   \left(4 t^3+3 t^2 u+3 t u^2+4 u^3\right)
\nonumber\\
&&-8 M^2 \left(5
   t^4+11 t^3 u+3 t^2 u^2+11 t u^3+5 u^4\right)+11 \left(4
   t^5+3 t^4 u+7 t^3 u^2+7 t^2 u^3+3 t u^4+4
   u^5\right)\bigg]
\nonumber\\
&&\bigg/\bigg[{9 M^5 t u (t+u)^3}\bigg].
\end{eqnarray}
The short distance coefficients of $^3P_J^{[8]}$ for $q\bar{q}{\rightarrow}{J/\psi}g$ subprocess were:
\begin{eqnarray}
&& \hspace{-0.9cm}\frac{F_{q\bar{q}}({}^3P_J^{[8]})}{m_c^4}=-\frac{1}{16\pi
s^2}\frac{1}{9}\frac{1}{4}\frac{(4\pi\alpha_s)^3}{N_c^2-1}\frac{640 \left(8 M^4 (t+u)-4 M^2 \left(t^2+4 t
   u+u^2\right)+3 \left(t^3+t^2 u+t
   u^2+u^3\right)\right)}{3 M^3 (t+u)^3
   \left(-M^2+t+u\right)}.
\end{eqnarray}
\begin{eqnarray}
&& \hspace{-0.9cm}\frac{G_{q\bar{q}}({}^3P_J^{[8]})}{m_c^6}=\frac{1}{16\pi
s^2}\frac{1}{9}\frac{1}{4}\frac{(4\pi\alpha_s)^3}{N_c^2-1}
\nonumber\\
&&256 \bigg[160 M^6 (t+u)-16 M^4 \left(5 t^2+17 t u+5
   u^2\right)+4 M^2 \left(t^3-11 t^2 u-11 t
   u^2+u^3\right)
\nonumber\\
&&+31 (t+u)^2 \left(t^2+u^2\right)\bigg]\bigg/\bigg[{3
   M^5 (t+u)^4 \left(-M^2+t+u\right)}\bigg].
\end{eqnarray}

The short distance coefficients of $^1S_0^{[8]}$ for $gq(\bar{q}){\rightarrow}{J/\psi}q(\bar{q})$ subprocess were
\begin{equation}
 \hspace{-0.9cm}\frac{F_{gq(\bar{q})}({}^1S_0^{[8]})}{m_c^2}=-\frac{1}{16\pi
s^2}\frac{1}{24}\frac{1}{4}\frac{(4\pi\alpha_s)^3}{N_c^2-1}\frac{160 \left(s^2+u^2\right)}{3 M (s+u)^2
   \left(-M^2+s+u\right)}.
\end{equation}
\begin{eqnarray}
 &&\hspace{-0.9cm}\frac{G_{gq(\bar{q})}({}^1S_0^{[8]})}{m_c^4}=\frac{1}{16\pi
s^2}\frac{1}{24}\frac{1}{4}\frac{(4\pi\alpha_s)^3}{N_c^2-1}\frac{320 \left(M^2 \left(11 s^3+23 s^2 u-s u^2+11
   u^3\right)-5 s \left(s^3+s^2 u+s
   u^2+u^3\right)\right)}{9 M^3 \left(M^2-s\right) (s+u)^3
   \left(M^2-s-u\right)}.
\end{eqnarray}
The short distance coefficients of $^3S_1^{[8]}$ for $gq(\bar{q}){\rightarrow}{J/\psi}q(\bar{q})$ subprocess were
\begin{eqnarray}
 \hspace{-0.9cm}\frac{F_{gq(\bar{q})}({}^3S_1^{[8]})}{m_c^2}=-\frac{1}{16\pi
s^2}\frac{1}{24}\frac{1}{4}\frac{(4\pi\alpha_s)^3}{N_c^2-1}\frac{1}{3}\frac{64 \left(4 s^2-s u+4 u^2\right) \left(2 M^4-2 M^2
   (s+u)+s^2+u^2\right)}{3 M^3 s u (s+u)^2}.
\end{eqnarray}
\begin{eqnarray}
&& \hspace{-0.9cm}\frac{G_{gq(\bar{q})}({}^3S_1^{[8]})}{m_c^4}=\frac{1}{16\pi
s^2}\frac{1}{24}\frac{1}{4}\frac{(4\pi\alpha_s)^3}{N_c^2-1}\frac{1}{3}
\nonumber\\
&&128 \bigg[2 M^6 \left(20 s^3+69 s^2 u-39 s u^2+20
   u^3\right)-2 M^4 \left(40 s^4+113 s^3 u+27 s^2 u^2+10 s
   u^3+20 u^4\right)
\nonumber\\
&&+M^2 \left(108 s^5+193 s^4 u+41 s^3
   u^2+225 s^2 u^3+s u^4+20 u^5\right)
\nonumber\\
&&-11 s \left(4 s^5+3
   s^4 u+7 s^3 u^2+7 s^2 u^3+3 s u^4+4 u^5\right)\bigg]\bigg/\bigg[{9
   M^5 s u \left(M^2-s\right) (s+u)^3}\bigg].
\end{eqnarray}
The short distance coefficients of $^3S_J^{[8]}$ for $gq(\bar{q}){\rightarrow}{J/\psi}q(\bar{q})$ subprocess were
\begin{equation}
 \hspace{-0.9cm}\frac{F_{gq(\bar{q})}({}^3P_J^{[8]})}{m_c^4}=\frac{1}{16\pi
s^2}\frac{1}{24}\frac{1}{4}\frac{(4\pi\alpha_s)^3}{N_c^2-1}\frac{640 \left(8 M^4 (s+u)-4 M^2 \left(s^2+4 s
   u+u^2\right)+3 \left(s^3+s^2 u+s
   u^2+u^3\right)\right)}{3 M^3 (s+u)^3
   \left(-M^2+s+u\right)}.
\end{equation}
\begin{eqnarray}
 &&\hspace{-0.9cm}\frac{G_{gq(\bar{q})}({}^3P_J^{[8]})}{m_c^6}=-\frac{1}{16\pi
s^2}\frac{1}{24}\frac{1}{4}\frac{(4\pi\alpha_s)^3}{N_c^2-1}
\nonumber\\
&&256 \bigg[8 M^6 \left(5 s^2+26 s u+25 u^2\right)+4
   M^4 \left(s^3-23 s^2 u-111 s u^2-19 u^3\right)
\nonumber\\
&&+M^2
   \left(57 s^4+226 s^3 u+166 s^2 u^2+58 s u^3+61
   u^4\right)-31 s (s+u)^2 \left(s^2+u^2\right)\bigg]
\nonumber\\
&&\bigg/\bigg[{3
   M^5 \left(M^2-s\right) (s+u)^4 \left(-M^2+s+u\right)}\bigg].
\end{eqnarray}


\end{document}